\documentclass[aps,pre,twocolumn,floatfix,groupedaddress,footinbib,notitlepage,showpacs,superscriptaddress,longbibliography]{revtex4-1}
\usepackage{times,bm,bbm,bbold,graphicx,graphics,amssymb,amsmath,amsthm,amsfonts,dsfont,hyperref,color}

\newcommand{\ket}[1]{|#1\rangle}
\newcommand{\bra}[1]{\langle#1|}
\newcommand{\braket}[2]{\langle #1|#2\rangle}
\newcommand{\ketbra}[2]{|#1\rangle\langle #2|}
\newcommand{\mrm}[1]{\mathrm{#1}}

\newcommand{\ignore}[1]{}

\newcommand{\be}{\begin{equation}}
\newcommand{\ee}{\end{equation}}

\let\oldsqrt\sqrt
\def\sqrt{\mathpalette\DHLhksqrt}
\def\DHLhksqrt#1#2{%
\setbox0=\hbox{$#1\oldsqrt{#2\,}$}\dimen0=\ht0
\advance\dimen0-0.2\ht0
\setbox2=\hbox{\vrule height\ht0 depth -\dimen0}%
{\box0\lower0.4pt\box2}}

\DeclareFontFamily{OT1}{pzc}{}
\DeclareFontShape{OT1}{pzc}{m}{it}%
              {<-> s * [1.25] pzcmi7t}{}
\DeclareMathAlphabet{\mathpzc}{OT1}{pzc}%
                                 {m}{it}

\DeclareMathOperator*{\Motimes}{\text{\raisebox{0.25ex}{\scalebox{0.8}{$\bigotimes$}}}}

\begin{document}

\title{Symmetry-induced fluctuation relations for dynamical observables irrespective of their behavior under time-reversal}

\author{Stefano Marcantoni}

\email{stefano.marcantoni@nottingham.ac.uk}
\affiliation{School of Physics \& Astronomy, University of Nottingham, Nottingham NG7 2RD, UK}
\affiliation{Centre for the Mathematics and Theoretical Physics of Quantum Non-equilibrium
Systems, University of Nottingham, Nottingham NG7 2RD, UK}

\author{Carlos P\'erez-Espigares}

\email{carlosperez@ugr.es}
\affiliation{Departamento de Electromagnetismo y F\'isica de la Materia, Universidad de Granada, Granada 18071, Spain}
\affiliation{Institute Carlos I for Theoretical and Computational Physics, Universidad de Granada, Granada 18071, Spain}

\author{Juan P. Garrahan}

\email{juan.garrahan@nottingham.ac.uk}
\affiliation{School of Physics \& Astronomy, University of Nottingham, Nottingham NG7 2RD, UK}
\affiliation{Centre for the Mathematics and Theoretical Physics of Quantum Non-equilibrium
Systems, University of Nottingham, Nottingham NG7 2RD, UK}

\begin{abstract}
We extend previous work to describe a class of fluctuation relations (FRs) that emerge as a consequence of symmetries at the level of stochastic trajectories in Markov chains. We prove that given such a symmetry, and for a suitable dynamical observable, it is \emph{always} possible to obtain a FR under a biased dynamics corresponding to the so-called generalized Doob transform. The general transformations of the dynamics that we consider go beyond time-reversal or spatial isometries, and an implication is the existence of FRs for observables irrespective of their behaviour under time-reversal, for example for time-symmetric observables rather than currents.
We further show how to deduce in the long-time limit these FRs from the symmetry properties of the generator of the dynamics. We illustrate our results with four examples that highlight the novel features of our work.
\end{abstract}

\maketitle

\section{Introduction}

Symmetries at the level of fluctuations or ``fluctuation relations" (FRs) that hold far from equilibrium are one of the most general results of nonequilibrium statistical mechanics. First discovered at the end of the last century, with the celebrated Gallavotti-Cohen fluctuation theorem \cite{evans93a, gallavotti95a, gallavotti95b} being the prominent example, fluctuation relations represent the macroscopic footprint of a microscopic symmetry breaking by constraining the probability distribution of time-integrated observables for systems away from equilibrium. Since then, a lot of theoretical work has been devoted to the study of fluctuation relations both in the classical and in the quantum domain \cite{jarzynski97b,kurchan98a,lebowitz99a,maes99a,crooks00a,hatano01a,collin05a,harris07a,andrieux07a,andrieux09a,perez-espigares12a,chetrite12a,ramezani18a,manzanog18a,gherardini18a,timpanaro19a}. For reviews see \cite{ritort08a,seifert12a,esposito09a,campisi11a}. 

Apart from the Gallavotti-Cohen fluctuation theorem - dealing with probabilities of an event and its time-reversal - other symmetries regarding spatial transformations, such as isometric fluctuation relations, have been unveiled in the last decade. This kind of relations were firstly introduced in the context of two-dimensional diffusive systems, by relating the probability of any pair of rotated currents \cite{hurtado11b} under some strong hypotheses that were subsequently removed \cite{perez-espigares16a}. The generalization to anisotropic systems \cite{villavicencio14a} helped to test experimentally their validity by measuring the velocity fluctuations of a self-propelled rod \cite{kumar15a} and those of hot Brownian swimmers \cite{falasco16a}. This also triggered some works on the emergence of FRs for static observables in equilibrium systems with broken symmetries \cite{lacoste14a,lacoste15a}.
Moreover, FRs for time-symmetric and activity-related quantities under involutions were discussed in \cite{maes06a,maes14a}.
More recently, a thermodynamic uncertainty relation \cite{barato15a,gingrich16a} has been derived for fluxes that satisfy an isometric FR \cite{vroylandt20a}. Although a spatial FR was introduced from a macroscopic perspective \cite{hurtado11b}, its microscopic derivation was  provided for Markovian stochastic systems in \cite{perez-espigares15a} under some assumptions on the dynamics.

Here we build on the results of Refs.~\cite{perez-espigares15a} and \cite{maes06a,maes14a} to generalise FRs that emerge as a consequence of symmetries in the dynamics. We do so in the framework of ``thermodynamics of trajectories'' 
\cite{eckmann85a,ruelle_2004,lecomte07c,garrahan09a,touchette09a,garrahan18a,Jack2019}, which extends the ensemble method of equilibrium statistical mechanics to dynamics. 
We show that given a dynamics which is symmetric under a certain transformation at the level of its trajectories, then a suitable observable can \emph{always} be found that defines a related dynamics satisfying a FR. This new dynamics is one whose trajectory ensemble is exponentially biased with respect to the original one, which is achieved by means of a generalized Doob transform \cite{Ahamed2006,Todorov2009,jack10a,chetrite15b,garrahan16a} that provides the optimal stochastic dynamics realizing a given fluctuation in the relevant observable. 
For long times, corresponding to the regime of large deviations \cite{touchette09a}, we show that from the symmetries of the generator it is possible to find the transformations which give rise to the FR.

The paper is structured as follows. In Sect.\ \ref{sec: StatTraj} we review the basic formalism to study the statistics of trajectories in continuous-time Markov chains. By means of this formalism, we present in Sect.\ \ref{sec: IsoFluc}  the FR introduced in \cite{perez-espigares15a} discussing its hypotheses  and showing how it can be generalized. In particular, we comment on the choice of the relevant observable and we point out that, given a symmetry of the original dynamics and a suitable observable, one can always obtain a FR by means of a proper conjugated dynamics through the generalized Doob transform. We further show how to obtain a FR from the symmetries of the generator. We also compare our findings with other results on time-symmetric observables already established in the literature. In Sect.\ \ref{sec: examples} we present four concrete examples which illustrate the novelty of our general results. Section \ref{sec: conclusion} gives our conclusions.

\section{Statistics of trajectories and generalized Doob transform}
\label{sec: StatTraj}

For concreteness we focus on dynamics described by continuous-time Markov chains. Central to our analysis will be the framework known as ``thermodynamics of trajectories'' 
\cite{eckmann85a,ruelle_2004,lecomte07c,garrahan09a,touchette09a,garrahan18a,Jack2019}
whereby the standard ensemble method of equilibrium statistical mechanics is extended to 
ensembles of trajectories of the dynamics. A trajectory $\omega_t$ up to time $t$ is fully characterized by a sequence of configurations of the system $x_0,x_1,x_2,\ldots,x_N$ together with the times of jump between them $t_1,t_2,\ldots,t_N$:
\begin{equation}
  \omega_t : x_0 \overset{t_1}{\longrightarrow} x_1 \overset{t_2}{\longrightarrow} x_2  \, \ldots \, x_{N-1}\overset{t_N}{\longrightarrow} x_N \,.
  \nonumber
\end{equation}
For simplicity we consider in the following systems with a finite number of configurations. 

The dynamics is determined by specifying the generator. This can be described using an operator formalism as, see e.g.\ \cite{schutz01a,garrahan18a}
$$
\partial_t \ket{P(t)}=\mathcal{L}\ket{P(t)}
$$
with probability vector $\ket{P(t)}=\sum_x P(x,t)\ket{x}$, where $P(x,t)=\braket{x}{P(t)}$ is the probability of being in configuration $x$ at time $t$ and the generator $\mathcal{L}$ reads
\begin{equation}
\mathcal{L} = \sum_{x,y\ne x} W_{x \to y} \ketbra{y}{x} - \sum_x R_x \ketbra{x}{x},
\label{gen0}
\end{equation}
with $\{ \ket{x}\}$ being an orthonormal basis of configurations, such that $\braket{x}{x'}=\delta_{x,x'}$. Here $W_{x\to y}$ are the jump rates between a pair of configurations $x$ and $y$, and $R_x=\sum_y W_{x\to y}$ is the escape rate from configuration $x$. The probability of a certain trajectory $\omega_t$ is then given by
\begin{equation}
    P(\omega_t) = \mathrm{e}^{-(t-t_N)R_{x_N}} W_{x_{N-1} \to x_N} \ldots \mathrm{e}^{-t_1R_{x_0}}  W_{x_0 \to x_1}P_{x_0},
      \nonumber
\end{equation}
where $P_{x_0}\equiv P(x_0,0)$ is the probability of being in $x_0$ at $t=0$.
In this context, an observable is a functional on the trajectory space. It is customary to distinguish between two types of observables \cite{garrahan09a}: type-A observables are related to the jumps occurring in the trajectory, while type-B observables are related to the time spent in each configuration. More explicitly, type-A observables are of the form
\begin{equation}
  A(\omega_t)= \sum_{x,y} Q_{x \to y}(\omega_t) \alpha_{x \to y},
    \label{typeAobs}
\end{equation}
where $Q_{x \to y}$ is the number of jumps (or ``flux'') from $x$ to $y$ in a trajectory $\omega_t$ and $\alpha_{x \to y}$ are real parameters accounting for the contribution to the observable $A(\omega_t)$ of each jump. For time-symmetric observables we have $\alpha_{x \to y}=\alpha_{y \to x}$. Instead, type-B observables are the time-integral of configurational functions,
\begin{equation}
  B(\omega_t)  = \int_0^t \mathrm{d}t' \, \beta[x(t')]\,
  \label{typeBobs}
\end{equation}
with $\beta$ being the quantity of interest evaluated in the configuration $x$ at time $t'$. 
A typical example of type-A observable is the {\em dynamical activity} \cite{lecomte07c,garrahan09a,garrahan18a,Maes2019}, namely the total number of jumps in a trajectory. This corresponds to taking $\alpha_{x \to y}=1$ for any pair of connected configurations $(x,y)$. An example of type-B observable is the time-integral of the  magnetization in the trajectory of a spin system.

The statistics of a stochastic observable $K(\omega_t)$ can be retrieved by computing $P(\omega_t)$ or alternatively from the moment generating function $Z(s)$, that reads
\begin{equation}
  Z(s)= \sum_{\omega_t} \mathrm{e}^{-s  K(\omega_t)} P(\omega_t) .
  \nonumber
\end{equation}
The above equation is as well the normalization factor of the exponentially biased distribution 
\begin{equation}
  P_s(\omega_t)= \frac{\mathrm{e}^{-s  K(\omega_t)} P(\omega_t)}{Z(s)},
  \label{Psensemble}
\end{equation}
known as the s-ensemble \cite{hedges09a}, which allows for the exploration of the rare events of interest through the parameter $s$.
Using the operator formalism it can be shown that the moment generating function can be computed as the following scalar product \cite{lecomte07c,garrahan09a,touchette09a,garrahan18a,Jack2019}, $Z(s) = \bra{-} \mathrm{e}^{t \mathcal{L}_s} \ket{P(0)}$, where $\bra{-}= \sum_x \bra{x}$ is the so-called flat state and the operator $\mathcal{L}_s$ is a tilted generator that reads
\begin{equation}
   \mathcal{L}_s =   \sum_{x,y\ne x} \mathrm{e}^{-s \alpha_{x \to y}} W_{x \to y} \ketbra{y}{x} - \sum_x R_x \ketbra{x}{x} ,
     \label{tiltAobs}
\end{equation}
for a type-A observable, or
\begin{equation}
   \mathcal{L}_s =   \sum_{x,y\ne x}  W_{x \to y} \ketbra{y}{x} - \sum_x \big( R_x + s\beta(x) \big)\ketbra{x}{x} ,
     \label{tiltBobs}
\end{equation}
for a type-B observable.

For large times the moment generating function satisfies a large deviation principle \cite{lecomte07c,garrahan09a,touchette09a,garrahan18a,Jack2019}
\begin{equation}
    Z(s) \approx \mathrm{e}^{t \theta(s)},
      \nonumber
\end{equation}
where $\theta(s)$ corresponds to the scaled cumulant generating function, which can be obtained as the largest eigenvalue of the tilted generator. At finite times the statistics of the selected observable depends on the full spectrum (and on the eigenvectors) of the tilted generator while at long times all the information concentrates in the largest eigenvalue. It is worth noting that the long-time average of the observable $K(\omega_t)$ in the s-ensemble \eqref{Psensemble} is given by 
\be
\frac{\langle K(\omega_t) \rangle_s}{t}=-\theta'(s)\, .
\ee
Unlike the original generator $\mathcal{L}$, corresponding to the case $s=0$, the tilted one is not a proper stochastic generator in the sense that it does not conserve probability, $\bra{-}\mathcal{L}_s \neq 0$. However, it is possible to construct a proper stochastic generator (in general time-dependent) such that rare trajectories of the original process are mapped into typical trajectories of the new one. This is realized through the generalized Doob transform that produces the time-dependent generator $\mathcal{L}^{\mrm{Doob}}_{t'} (s)$ \cite{jack10a,chetrite15b,garrahan16a}
\begin{equation}\label{tdoob}
\mathcal{L}^{\mrm{Doob}}_{t'} (s) = G_{t'} \mathcal{L}_s G_{t'}^{-1} - \partial_{t'} \log(Z(s)) + (\partial_{t'} G_{t'}) G_{t'}^{-1} 
\nonumber
\end{equation}
by means of the gauge transformation $G_{t'}$
\begin{equation}
    G_{t'}= \sum_x \frac{\langle - | \mathrm{e}^{(t-t')\mathcal{L}_s} | x \rangle}{\langle - | \mathrm{e}^{(t-t')\mathcal{L}_s} | x_0 \rangle} \ketbra{x}{x} ,
    \nonumber
\end{equation}
where $t$ is the final time. For asymptotically long times $t-t' \gg 1$ the exponential operator is well approximated as $\mathrm{e}^{(t-t')\mathcal{L}_s} \simeq \mathrm{e}^{(t-t')\theta(s)} \ketbra{r_0}{\ell_0} $, where $\ket{r_0}$ and $\bra{\ell_0}$ are the right and left eigenvectors of $\mathcal{L}_s$ corresponding to the largest eigenvalue $\theta(s)$, namely $\bra{\ell_0} \mathcal{L}_s = \theta(s) \bra{\ell_0}$ and $\mathcal{L}_s \ket{r_0}= \theta(s) \ket{r_0}$, which are normalized as $\braket{\ell_0}{r_0}=\braket{-}{r_0}=1$. Therefore, the gauge transformation becomes time-independent and reads $G_\infty = \braket{\ell_0}{x_0}^{-1}\sum_x \braket{\ell_0}{x} \ketbra{x}{x}$ so that in the end one has the following generator for long times \cite{jack10a,chetrite15b,garrahan16a}
\begin{equation}
   \mathcal{L}^{\mrm{Doob}}_{\infty} (s)= G_\infty \mathcal{L}_s G_\infty^{-1} -\theta(s) . 
   \nonumber
\end{equation}
Moreover, defining the matrix $L_s$ as the matrix connecting the left eigenvector and the flat state $\bra{\ell_0}=\bra{-}L_s$ we get that $L_s=G_\infty \braket{\ell_0}{x_0}$. Thus assuming $L_s$ is invertible, we can write the long-time Doob generator $\mathcal{L}^{\mrm{Doob}}(s)\equiv \mathcal{L}^{\mrm{Doob}}_{\infty} (s) $ as
\begin{equation}
\mathcal{L}^{\mrm{Doob}}(s) = L_s \mathcal{L}_s L_s^{-1} -\theta(s) , 
\label{Dooblong}
\end{equation}
which corresponds to a proper stochastic generator such that  $ \bra{-}  \mathcal{L}^{\mrm{Doob}}(s)=0$.
One can show that the time-dependent Doob generator \eqref{tdoob} describes an ensemble of stochastic trajectories with probability distribution given by \eqref{Psensemble}, i.e. that is exponentially biased with respect to the ensemble generated by the original dynamics (see assumption 3 in the next section) \cite{jack10a,chetrite13a,chetrite15b}. The time-independent generator $\mathcal{L}^{\mrm{Doob}}(s)$ generates instead $P_s(\omega_t)$ in the long-time limit.
In the examples of Sect.\ \ref{sec: examples} we will mainly use the time-independent Doob generator \eqref{Dooblong} as constructed above and comment on the time-dependent case in the second example.

\section{Fluctuation relation}
\label{sec: IsoFluc}

Fluctuation relations other than Gallavotti-Cohen such as FRs associated with spatial transformations were firstly introduced in the context of diffusive systems \cite{hurtado11b}. A derivation from the microscopic Markovian dynamics of this kind of FR was proved in \cite{perez-espigares15a} by means of three assumptions:
\begin{enumerate}
\item There is a bijection $\mathcal{R}$ in the {\em space of trajectories} such that $P_0(\mathcal{R}\omega_t)= P_0 (\omega_t)$,
\item There is an observable (maybe vectorial) $\underline{K}(\omega_t)$ such that $\underline{K}(\mathcal{R} \omega_t)= U \cdot \underline{K}(\omega_t)$ for some matrix $U$ (independent of $\omega_t$),
\item A modified dynamics exists such that the probability of a certain trajectory in this new dynamics is related to the probability under the original dynamics as follows
\begin{equation} 
P_{\underline{E}}(\omega_t)= \frac{\mathrm{e}^{-\underline{E}^T\cdot \underline{K}(\omega_t)}P_0(\omega_t)}{Z_0(\underline{E})},
\label{3assump}
\end{equation}
where $\underline{E}$ is a field that breaks the initial symmetry and $Z_0(\underline{E}) = \sum_{\omega_t} \mathrm{e}^{-\underline{E}^T\cdot \underline{K}(\omega_t)} P_0(\omega_t)$ is the normalization.
\end{enumerate}
Here and in the following, column vectors are indicated by $\underline{v}$ and row vectors $\underline{v}^T$ (with $T$ denoting transposition), while the dot $\cdot$ is the usual product of matrices.
The FR is then expressed as a symmetry of the moment generating function $Z_{\underline{E}}(\underline{\lambda})$ defined as follows
\begin{equation}
Z_{\underline{E}}(\underline{\lambda})= \sum_{\omega_t} \mathrm{e}^{-\underline{\lambda}^T\cdot \underline{K}(\omega_t)} P_{\underline{E}}(\omega_t), 
\nonumber
\end{equation}
describing the statistics of the stochastic observable $\underline{K}$ in the modified dynamics. In particular, it turns out that
\begin{equation}\label{FR}
Z_{\underline{E}}(\underline{\lambda})= Z_{\underline{E}}[(U^{-1})^T \cdot (\underline{\lambda} + \underline{E}) - \underline{E}].
\end{equation}
The proof of this result is quite straightforward. Indeed, by means of the three assumptions presented, one can write the following chain of equalities
\begin{align}
Z_{\underline{E}}(\underline{\lambda}) &= \sum_{\omega_t} \mathrm{e}^{-\underline{\lambda}^T\cdot \underline{K}(\omega_t)} P_{\underline{E}}(\omega_t) \nonumber \\
&\overset{(3)}{=} \sum_{\omega_t} \mathrm{e}^{-\underline{\lambda}^T \cdot \underline{K}(\omega_t)} \frac{\mathrm{e}^{-\underline{E}^T\cdot \underline{K}(\omega_t)}P_0(\omega_t)}{Z_0(\underline{E})}  \nonumber \\ 
& \overset{(1)}{=} \sum_{\omega_t} \mathrm{e}^{-(\underline{\lambda}+\underline{E})^T \cdot \underline{K}(\omega_t)} \frac{P_0(\mathcal{R}\omega_t)}{Z_0(\underline{E})} \nonumber \\ 
&=\sum_{\omega_t} \mathrm{e}^{-(\underline{\lambda}+\underline{E})^T\cdot \underline{K}(\mathcal{R}^{-1}\omega_t)} \frac{P_0(\omega_t)}{Z_0(\underline{E})} \nonumber \\ 
&\overset{(3)}{=} \sum_{\omega_t} \mathrm{e}^{-(\underline{\lambda}+\underline{E})^T\cdot \underline{K}(\mathcal{R}^{-1}\omega_t)} \mathrm{e}^{\underline{E}^T\cdot \underline{K}(\omega_t)} P_{\underline{E}}(\omega_t) \nonumber \\
&\overset{(2)}{=} \sum_{\omega_t} \mathrm{e}^{-(\underline{\lambda}+\underline{E})^T\cdot U^{-1}\cdot \underline{K}(\omega_t)+\underline{E}^T\cdot \underline{K}(\omega_t)} P_{\underline{E}}(\omega_t) \nonumber \\
&= \sum_{\omega_t} \mathrm{e}^{-(\underline{\lambda}')^T\cdot \underline{K}(\omega_t)} P_{\underline{E}}(\omega_t) = Z_{\underline{E}}(\underline{\lambda}'),
\end{align}
where $\underline{\lambda}'= (U^{-1})^T \cdot (\underline{\lambda} + \underline{E}) - \underline{E}$ and the numbers parenthesis are used to clarify the role of each assumption. Remarkably, this relation is true at any finite time $t$. By looking at the behavior for asymptotically long times, one can also find out a symmetry relation at the level of the scaled cumulant generating function $\theta(\underline{\lambda})$. Indeed, since $Z_{\underline{E}}(\underline{\lambda})\sim \mathrm{e}^{t\theta_{\underline{E}}(\underline{\lambda})}$ for long times, it turns out that 
\begin{equation}
\theta_{\underline{E}}(\underline{\lambda})= \theta_{\underline{E}
}[(U^{-1})^T \cdot (\underline{\lambda} + \underline{E}) - \underline{E}].
\label{FRlongt}
\end{equation}

Our first contribution is to notice that the third assumption \eqref{3assump} is not an assumption, in the sense that, given a symmetry of $P_0(\omega_t)$ and a certain observable $\underline{K}(\omega_t)$, a dynamics satisfying Eq.\eqref{3assump} always exists.
This dynamics is provided by the generalized \emph{Doob transform} \cite{Ahamed2006,Todorov2009,jack10a,chetrite15b,garrahan16a}, briefly presented in the previous section, where the parameter $s$ (that can be vectorial) has the role of the external field $\underline{E}$. Thus, the Doob transform generates the ensemble of stochastic trajectories with probability $P_{\underline{E}}(\omega_t)$. Therefore, the fluctuation relation given in \eqref{FR} can be always found given the constant field $\underline{E}$, which biases the statistics of trajectories $P_0(\omega_t)$ and which breaks its symmetry property.  
This relation includes as special cases previously known results. For instance, if we choose the bijection $\mathcal{R}$ to be the time-reversal and the observable $K(\omega_t)$ to be a current (anti-symmetric under time reversal so that $U$ amounts to a minus sign) we recover the celebrated Gallavotti-Cohen relation for stochastic processes (see e.g. \cite{harris07a}). However, our result is more general inasmuch it deals also with transformations different from time-reversal, like spatial rotations and translations, and observables different from currents as for instance time-symmetric ones.

In a couple of works \cite{lacoste14a, lacoste15a} a similar fluctuation relation was proved, by comparing the equilibrium Gibbs distributions relative to a symmetric Hamiltonian and a modified one where a field breaks the symmetry. 
Our work can be considered as a result along the same lines, where equilibrium ensembles in configuration space are replaced by dynamical ensembles in trajectory space. Also, our findings apply to equilibrium stochastic dynamics (when detailed balance is satisfied) as well as out-of-equilibrium.
Active fluctuation symmetries are also discussed in the literature \cite{maes14a} pointing out that time-symmetric observables can also obey fluctuation relations \cite{maes06a}. In that context however, the analysis was limited to the study of involutions in the trajectory space. In this work instead we never use the assumption that the transformation $\mathcal{R}$ is an involution.

Following \cite{maes14a}, we can use the previous framework also to derive another fluctuation relation for a generic observable $f(\omega_t)$ in the modified dynamics
\begin{align}
\langle f(\mathcal{R}\omega_t) \rangle_{\underline{E}} &= \sum_{\omega_t} f(\mathcal{R}\omega_t) P_{\underline{E}}(\omega_t) \nonumber \\
&= \sum_{\omega_t} f(\mathcal{R}\omega_t) \frac{\mathrm{e}^{-\underline{E}^T\cdot \underline{K}(\omega_t)}P_0(\omega_t)}{Z_0(\underline{E})} \nonumber \\
&= \sum_{\omega_t} f(\omega_t) \frac{\mathrm{e}^{-\underline{E}^T\cdot U^{-1} \underline{K}(\omega_t)}P_0(\omega_t)}{Z_0(\underline{E})}  \nonumber \\
&= \sum_{\omega_t} f(\omega_t) \mathrm{e}^{-(\underline{E}^T\cdot U^{-1} - \underline{E}^T)\cdot \underline{K}(\omega_t)}  P_{\underline{E}}(\omega_t)\nonumber \\
&=\langle f(\omega_t)\,\mathrm{e}^{-(\underline{E}^T\cdot U^{-1} - \underline{E}^T)\cdot \underline{K}(\omega_t)}   \rangle_{\underline{E}} .\nonumber
\end{align} 
Considering the constant function $f(\omega_t)=1$ one obtains as a consequence a Jarzynski-like fluctuation relation
\begin{equation}\label{Jarz}
\Big\langle \mathrm{e}^{(\underline{E}^T - \underline{E}^T\cdot U^{-1})\cdot \underline{K}(\omega_t)}  \Big\rangle_{\underline{E}}  =1,
\end{equation}
and applying Jensen's inequality this in turn gives a constraint on the average of the exponent
\begin{equation}
\Big\langle (\underline{E}^T -\underline{E}^T\cdot U^{-1} )\cdot \underline{K}(\omega_t) \Big\rangle_{\underline{E}} \leq 0,
\end{equation}
or equivalently
\begin{equation}
 {\underline{E}}^T \cdot \Big\langle \underline{K}(\mathcal{R}\omega_t) \Big\rangle_{\underline{E}} \leq  {\underline{E}}^T \cdot  \Big\langle \underline{K}(\omega_t) \Big\rangle_{\underline{E}} .
\end{equation}
This inequality provides a constraint on the average of the observable $\underline{K}$ in the modified dynamics with respect to the same observable evaluated on the transformed trajectory.

The results presented so far are valid for general bijections in the trajectory space. In the following, for the sake of convenience, we restrict the discussion to transformations at the configuration level.

\subsection{Choice of the observable}
\label{subsec: obs}

We now consider the choice of the, in general vectorial, observable that satisfies the second assumption above. In particular, we show that it is always possible to find such an observable provided its dimension is sufficiently high and the transformation is actually a transformation in the configuration space. 
Consider a type-A observable $\underline{A}$, as defined in \eqref{typeAobs}, namely an observable related to the jumps between two configurations, whose components $A_a$ are written as follows
\begin{equation}
A_a(\omega_t)= \sum_{x,y} Q_{x \to y}(\omega_t) \alpha^a_{x \to y},
\end{equation}
given the total number of jumps from $x$ to $y$ in the trajectory $\omega_t$, $Q_{x \to y}(\omega_t)$, and a set of real parameters $\alpha^a_{x \to y}$. In a system with $D$ possible configurations, in continuous time, the maximum number of allowed jumps is $D(D-1)$, in the case of a fully connected problem. Therefore, a generic observable $\underline{A}$ belongs to a $D(D-1)$ dimensional vector space, being a linear combination of the different number of jumps $Q_{x \to y}$ with real coefficients. 

Consider now a bijective transformation $R$ acting on the configuration space. This in turn induces a map $\mathcal{R}$ in the trajectory space given by
\begin{eqnarray}
\nonumber
& \omega_t: x_1 \to x_2 \to \ldots \to x_N \\ \nonumber
&\mathcal{R}\big\downarrow \\ 
&\mathcal{R} \omega_t: R x_1 \to R x_2 \to \ldots \to R x_N. 
\label{Rconf}
\end{eqnarray}
As a consequence, the number of jumps between two configurations $x$ and $y$ in the original trajectory $\omega_t$ equals the number of jumps between the transformed configurations $R x$ and $R y$ in the transformed trajectory $\mathcal{R} \omega_t$, thus
\begin{equation}
Q_{Rx \to Ry} (\mathcal{R}\omega_t) = Q_{x \to y}(\omega_t).
\label{qRtilde}
\end{equation}
The observable $\underline{A}$ in the modified trajectory then reads
\begin{align}\label{Ktilde1}
A_a(\mathcal{R}\omega_t) &=\sum_{x,y}Q_{x \to y}(\mathcal{R}\omega_t) \alpha^a_{x \to y} \nonumber \\
&= \sum_{x,y}Q_{R^{-1}x \to R^{-1}y}(\omega_t) \alpha^a_{x \to y}  \nonumber \\
&= \sum_{x,y}Q_{x \to y}(\omega_t) \alpha^a_{Rx \to Ry} = \widetilde{A}_a(\omega_t),
\end{align}
where the first step is a consequence of \eqref{qRtilde} and the second one is just a change of variable. We want to find the linear transformation $U$ that relates the original observable to $\widetilde{\underline{A}}(\omega_t)= U \cdot \underline{A}(\omega_t)$ , with components
\begin{equation}\label{Ktilde2}
\widetilde{A}_a(\omega_t) = \sum_b U_{a b} A_b(\omega_t) = \sum_{x,y} Q_{x \to y} (\omega_t) \sum_b U_{a b} \alpha^b_{x \to y}.
\end{equation}
By comparing \eqref{Ktilde1} and \eqref{Ktilde2} one finds that 
\begin{equation}
U\cdot \underline{\alpha}_{x\to y}=\underline{\alpha}_{Rx\to Ry}\, , 
\label{alphatransf}
\end{equation}
i.e. that the $D^2(D-1)^2$ elements of the matrix $U$ have to satisfy the system of linear equations $\sum_b U_{a b} \alpha^b_{x \to y} = \alpha^a_{Rx \to Ry}$, for any pair $x,y$. These are in principle $D^2(D-1)^2$ equations so that the system should allow for a solution. In order to better understand the condition for a unique solution we concentrate for the moment on the simplest case of a $2D$ configuration space. Therefore, we consider a system with two possible configurations $x,y$ so that just two different jumps ($D(D-1)=2$) are possible $x \to y$ and $y \to x$. Moreover one can have just two ($D!=2$) different bijections $R_1$ and $R_2$, where 
\begin{align}
&R_1 x = x, \quad R_1 y = y, \nonumber \\ 
&R_2 x = y, \quad R_2 y = x. \nonumber
\end{align} 
The maximun number of parameters $\alpha$ is $D^2(D-1)^2=4$, indeed one has the four real parameters
\begin{equation}
\alpha^{1}_{x \to y}, \alpha^{1}_{y \to x}, \alpha^{2}_{x \to y}, \alpha^{2}_{y \to x}, \nonumber
\end{equation}
that allow us to write the equations for the matrix elements of $U$ as
\begin{equation}
\underbrace{\begin{pmatrix}
\alpha^{1}_{x \to y} &\alpha^{2}_{x \to y} &0 &0 \\
\alpha^{1}_{y \to x} &\alpha^{2}_{y \to x} &0 &0 \\
0 &0 &\alpha^{1}_{x \to y} &\alpha^{2}_{x \to y} \\
0 &0 &\alpha^{1}_{y \to x} &\alpha^{2}_{y \to x} \\
\end{pmatrix} }_{\text{\normalsize{ $M$ }}} \nonumber
\begin{pmatrix}
u_{11} \\
u_{12} \\
u_{21} \\
u_{22}
\end{pmatrix} =
\begin{pmatrix}
\alpha^{1}_{Rx \to Ry} \\
\alpha^{1}_{Ry \to Rx} \\
\alpha^{2}_{Rx \to Ry} \\
\alpha^{2}_{Ry \to Rx}
\end{pmatrix} .
\end{equation}
Therefore, the solution is unique if and only if the matrix $M$ has nonzero determinant, that in turn, due to the block-diagonal structure of $M$, corresponds to have $\mathrm{det}(\alpha)\neq 0$, where the matrix $\alpha$ is
\begin{equation}
\alpha = 
\begin{pmatrix}
\alpha^{1}_{x \to y} &\alpha^{2}_{x \to y} \\
\alpha^{1}_{y \to x} &\alpha^{2}_{y \to x}
\end{pmatrix}.\nonumber
\end{equation}
The nonzero determinant implies the two components of the vectorial observable are indeed linearly independent. Otherwise one could recast them in a scalar observable and the dimensional argument would not work any more. The same reasoning holds true in higher dimensions so that it is always possible to construct a suitable observable (even though maybe not so relevant from a physical point of view) so that the assumption number $2$ is satisfied in a fully connected problem. If the system is not fully connected, one can restrict the previous discussion to the number of allowed jumps (strictly less than $D(D-1)$) and everything applies in the same way, thus implying the validity of assumption number $2$. In this case, one has to be careful with the choice of the bijection $R$ in configuration space. Indeed, only those bijections $R$ that preserve the set of allowed jumps induce a bijection $\mathcal{R}$ on the trajectories of the system. This can be easily seen considering a totally asymmetric random walk on a ring (for simplicity let us just consider $4$ sites)
$$
\begin{matrix}
1 &\rightarrow  &2 \\ 
\uparrow &{} &\downarrow \\
4 &\leftarrow &3
\end{matrix}
$$
A transformation $R$ such that $R1=3, R2=2$ and $R3=1$ does not induce a bijection $\mathcal{R}$ in the trajectory space of the system because for instance $\omega_t: 1 \to 2 \to 3  $ should be mapped into $\widetilde{\omega}_t: 3 \to 2 \to 1$ that is not allowed. Instead, a transformation like  $Rx=x+1 (\text{MOD}\, 4)$, preserves the set of allowed jumps and is therefore acceptable. 
 This will be the situation discussed in the examples of Section \ref{sec: examples}. More precisely, in those examples we will show that it is usually possible and more interesting to find low dimensional observables satisfying the assumption number $2$. Further comments on the construction of low dimensional observables can be found in Appendix \ref{app: observables}.

\subsection{FR from the symmetries of the generator}
The FR \eqref{FR} presented above is very compelling as it relates the probability of different fluctuations \emph{at all times} from a symmetry of the probability of trajectories, $P_0(\omega_t)=P_0(\mathcal{R}\omega_t)$. However, for bijective transformation acting on configurations---as in \eqref{Rconf}---the symmetry $P_0(\omega_t)=P_0(\mathcal{R}\omega_t)$ holds when both the transition rates and the probability of the initial state are symmetric under the transformation, namely $W_{x \to y}= W_{Rx \to Ry}$ and $P_{x_0}=P_{Rx_0}$. This might be something difficult to have, since we should prepare the system in an initial symmetric state. Yet, we show in this section that we can derive a FR for \emph{long times}, i.e. $\theta_{\underline{E}}(\underline{\lambda})=\theta_{\underline{E}}[(U^{-1})^T\cdot (\lambda + \underline{E})-\underline{\lambda}]$, just from the symmetries of the dynamical generator \eqref{gen0} --so that $W_{x \to y}= W_{Rx \to Ry}$--, without caring about the symmetries of the initial state.

We thus start by assuming that the original generator \eqref{gen0} has a certain symmetry under the transformation $R$, described by the operator $V$ such that $V\ket{x}=\ket{Rx}$,
\begin{equation}
\mathcal{L} = V \mathcal{L} V^T,
\label{tilsymmrel}
\end{equation}
that in turn implies $W_{x \to y}= W_{Rx \to Ry}$ for any pair $(x,y)$. From this we now prove the following similarity transformation for the tilted generator
\begin{equation}\label{flucgen}
 \mathcal{L}_{\lambda} = V \mathcal{L}_{(U^{-1})^T \cdot \underline{\lambda}} V^T . 
\end{equation}
Since the tilted generator with respect to a type-A observable is
\begin{equation}
\mathcal{L}_{\underline{\lambda}} = \sum_{x,y\ne x} \mathrm{e}^{-\underline{\lambda}^T\cdot \underline{\alpha}_{x \to y}} W_{x \to y} \ketbra{y}{x} - \sum_x R_x \ketbra{x}{x},
\nonumber
\end{equation}
the transformed one thus reads
\begin{equation}
V \mathcal{L}_{\underline{\lambda}} V^{T}=\!\!\! \sum_{x,y\ne x} \mathrm{e}^{-\underline{\lambda}^T\cdot \underline{\alpha}_{x \to y}} W_{x \to y} \ketbra{Ry}{Rx} - \sum_x R_x \ketbra{Rx}{Rx}. 
\nonumber
\end{equation}
By applying a change of variable and using the fact, as shown in \eqref{alphatransf}, that the parameters $\underline{\alpha}_{x \to y}$ transform according to
\begin{equation}
 \underline{\alpha}_{R^{-1}x \to R^{-1}y} = U^{-1} \cdot \underline{\alpha}_{x \to y}  
 \nonumber
\end{equation}
the relation \eqref{flucgen} follows immediately. Then, denoting $L_{\underline{s}}$ for the diagonal matrix whose entries corresponds to the left eigenvector associated with $\theta(\underline{s})$, and exploiting the relation between the tilted generator and the Doob one for long times \cite{jack10a,carollo18b}
\begin{equation}
 \mathcal{L}^{\mrm{Doob}}_{\lambda}(\underline{s}) = L_{\underline{s}} \mathcal{L}_{\underline{\lambda} + \underline{s}} L_{\underline{s}}^{-1} - \theta(\underline{s})   ,
 \label{tiltedDoob}
\end{equation}
we also find a symmetry relation at the level of the tilted Doob generator \eqref{tiltedDoob}: such generator describes the statistics of the observable in the Doob dynamics and satisfies the following symmetry relation,
\begin{equation}
 \mathcal{L}^{\mrm{Doob}}_{\lambda}(\underline{s}) = A_{\underline{s}} \mathcal{L}^{\mrm{Doob}}_{(U^{-1})^T\cdot (\underline{\lambda}+\underline{s})-\underline{s}}(\underline{s}) A_{\underline{s}}^{-1}
 \label{doobsymm}
\end{equation}
where $A_{\underline{s}} = L_{\underline{s}} V L_{\underline{s}}^{-1}$. Identifying $\underline{s}$ with the field $\underline{E}$ we see that \eqref{doobsymm} implies the FR \eqref{FRlongt}, 
\begin{equation}
\theta_{\underline{E}}(\underline{\lambda})=\theta_{\underline{E}}[(U^{-1})^T\cdot (\underline{\lambda} + \underline{E})-\underline{\lambda}].
\label{FRlong}
\end{equation}
The same result can be derived for type-B observables \eqref{typeBobs}, for which the tilted generator is given by \eqref{tiltBobs}.

We thus have demonstrated that from the symmetries of the generator for bijective transformations $R$ acting on configurations such that $\underline{K}(\mathcal{R}\omega_t)=U\cdot \underline{K}(\omega_t)$, the FR \eqref{FRlong} is derived.

\section{Examples}
\label{sec: examples}
We now consider four different examples of increasing complexity to illustrate the 
general FRs obtained above. The first two examples are analytically solvable and deal with a single particle hopping on a ring. The first one shows that a fluctuation relation can exist for an activity-like observable (symmetric under time-reversal) provided it can take both positive and negative values. The second example deals with a two-dimensional (time-symmetric) observable that only has positive entries. Indeed, in this case, the increased dimensionality is sufficient to provide the symmetry of the scaled cumulant generating function. The other two examples involve many-body dynamics and are related to the fluctuation of the magnetization, a type-B observable. In particular, we show that a fluctuation relation holds true for the time-integrated magnetization in a Glauber-Ising dynamics modified with a transverse field. By looking at the generator, we also discuss the same observable in the context of a three-state Potts model.

\subsection{Time-symmetric observable for an asymmetric random walk}
\begin{figure}[t]
\centering
\includegraphics[scale=0.37]{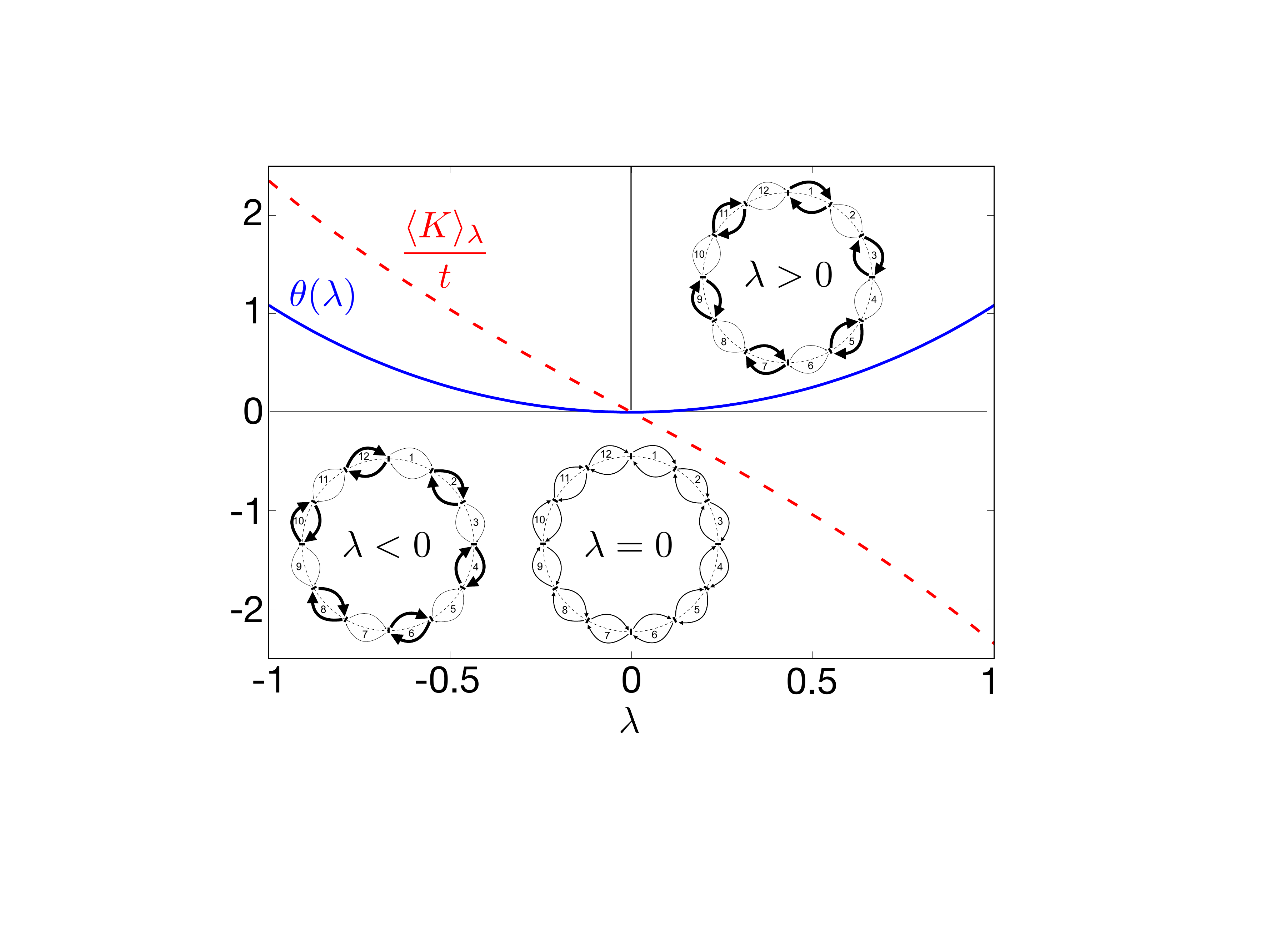}
\caption{{\bf Random walk on a ring}. Scaled cumulant generating function \eqref{scgfexA} (solid blue line) with $\gamma_1=\gamma_2=1$ for the observable $K$, namely the difference between the number of jumps in even and odd bonds, together with $\langle K \rangle_{\lambda}/t=-\theta'(\lambda)$ (red dashed line). Insets: Sketch of the Doob dynamics of the system \eqref{DoobexA} for $s=\lambda$ and different values of $\lambda$. Thicker arrows correspond to larger transition rates ($e^{|\lambda|}$) while thinner arrows correspond to smaller transition rates ($e^{-|\lambda|}$).}
\label{Fig1}
\end{figure}

Consider a particle performing an asymmetric random walk on a ring of $L$ sites. The particle jumps to the right with rate $\gamma_1$ and to the left with rate $\gamma_2$. The net number of jumps in a given trajectory corresponds to the time-integrated current, while the total number of jumps is the activity. However, in order to illustrate the FR derived above we focus on a time-extensive observable that is time-symmetric, but with the possibility to take positive and negative values. 
Therefore the observable we choose is 
$$
K(\omega_t)=K_{\mrm{even}}(\omega_t)-K_{\mrm{odd}}(\omega_t), 
$$
namely the difference between the number of jumps in even and odd bonds in a given trajectory. The mean stationary value of this observable is zero, since there is no asymmetry between bonds as the hopping rates are the same for any bond. We choose $L$ even for convenience so that we have the same number of even and odd bonds. 
The exponentially tilted generator of the process thus reads {\small
\begin{align*}
&\mathcal{L}_\lambda = \gamma_1 \mathrm{e}^{-\lambda} \sum_{x\, \mrm{even}} \ketbra{x+1}{x} + \gamma_2 \mathrm{e}^{-\lambda} \sum_{x\, \mrm{odd}} \ketbra{x-1}{x} +\nonumber \\
& + \gamma_1 \mathrm{e}^{\lambda} \sum_{x\, \mrm{odd}} \ketbra{x+1}{x} + \gamma_2 \mathrm{e}^{\lambda} \sum_{x\, \mrm{even}} \ketbra{x-1}{x} - (\gamma_1 + \gamma_2)\mathbbm{1} \,
\end{align*}}
where $\ket{x}$ is the configuration in which the particle is at site $x\in \{1,2,...,L\}$. This describes a situation where four kinds of jump are weighted differently, namely, apart from distinguishing clockwise and counterclockwise jumps as in the original process, the rate depends on the kind of bond being even or odd. From $\mathcal{L}_\lambda$ we see that positive values of $\lambda$ bias the dynamics towards a negative value of $K$, by enhancing the number of jumps in odd bonds, while negative values of $\lambda$ do the opposite (see Fig.~\ref{Fig1}, which is explained below). 
This generator can be diagonalized exactly. The eigenvalues $\xi_q$ satisfy the relation
\begin{equation*}
\left( \xi_q + \gamma_1 + \gamma_2 \right)^2 = 2\gamma_1 \gamma_2 \cosh(2\lambda) + \gamma_1^2 \mathrm{e}^{-2iq} + \gamma_2^2 \mathrm{e}^{2iq},
\end{equation*}
where $q = \frac{2\pi n}{L}$ with $n \in \{0,1, \ldots ,L-1\}$. The right eigenvectors can be written as {\small
\begin{equation*}
\ket{r_q(\lambda)}= \sum_{m=1}^{L/2} \left[ \mathrm{e}^{i2mq} c_{2m} \ket{2m} + \mathrm{e}^{i(2m-1)q} c_{2m-1}^q \ket{2m-1} \right],
\end{equation*} }
where $c_{2m}=c$ is a constant $\forall m$ and the odd coefficients $c_{2m-1}^q$ read
\begin{equation}
c_{2m-1}^q = c \frac{\gamma_1 \mathrm{e}^{-\lambda} \mathrm{e}^{-iq} + \gamma_2 \mathrm{e}^{\lambda}\mathrm{e}^{iq}}{\left( 2\gamma_1 \gamma_2 \cosh(2\lambda) + \gamma_2^2 \mathrm{e}^{2iq} + \gamma_1^2 \mathrm{e}^{-2iq}\right)^{1/2}}.
\end{equation}
The left eigenvectors are obtained by exchanging $q$ with $-q$ and $\gamma_1$ with $\gamma_2$. As a result of the normalization conditions $\braket{l_{q}}{r_{q'}}= \delta_{q q'}$ and $\braket{-}{r_0}= 1$ one finds that $c=1/L$. As a first check we can see that the scaled cumulant generating function
\begin{equation}
\theta(\lambda)= -(\gamma_1 + \gamma_2) + \sqrt{\gamma_1^2 + \gamma_2^2 + 2\gamma_1 \gamma_2 \cosh(2\lambda)} 
\label{scgfexA}
\end{equation}
is vanishing for $\lambda=0$ and satisfies the symmetry property $\theta(\lambda)= \theta(-\lambda)$. This symmetry is displayed in Fig.~\ref{Fig1}, where we show $\theta(\lambda)$ together $\langle K \rangle_{\lambda}/t$ for $\gamma_1=\gamma_2=1$. This is indeed the expected behaviour due to the properties of the original generator $\mathcal{L}_0$ that is symmetric under the shift of one site $\mathcal{L}_0=V\mathcal{L}_0 V^T$, $V \ket{x} = \ket{x+1}$, and due to the chosen observable that instead changes sign under the same transformation ($U=-1$). The same is true for a shift of any odd number of sites. 

We can now perform the Doob transform in order to find a proper stochastic generator where the initial symmetry is explicitly broken {\small
\begin{align}
&\mathcal{L}^{\mrm{Doob}}(s) = \gamma_1 \mathrm{e}^{-s} \alpha(s) \sum_{x\, \mrm{even}} \ketbra{x+1}{x} + \nonumber\\ 
& +\gamma_2 \mathrm{e}^{-s} \alpha(s)^{-1} \sum_{x\, \mrm{odd}} \ketbra{x-1}{x} + \gamma_1 \mathrm{e}^{s} \alpha(s)^{-1} \sum_{x\, \mrm{odd}} \ketbra{x+1}{x}+ \nonumber \\
& + \gamma_2 \mathrm{e}^{s} \alpha(s) \sum_{x\, \mrm{even}} \ketbra{x-1}{x}  - \sqrt{\gamma_1^2 + \gamma_2^2 + 2\gamma_1 \gamma_2 \cosh(2s)}\mathbbm{1},
\label{DoobexA}
\end{align} }
with $\alpha(s)$ being the following ratio 
\begin{equation}
\alpha(s)= \frac{\gamma_1 \mathrm{e}^{s}+ \gamma_2 \mathrm{e}^{-s}}{\sqrt{\gamma_1^2 + \gamma_2^2 + 2\gamma_1 \gamma_2 \cosh(2s)}}.
\end{equation}
Notice that for $\gamma_1=\gamma_2=\gamma$ we get $\alpha(s)=1$, so that the rates in the biased stochastic dynamics given by \eqref{DoobexA} become $\gamma e^{-s}$ for clock- and counter-clockwise jumps over even bonds and $\gamma e^{s}$, also in both directions, for odd bonds. This has been sketched in the insets to Fig.~\ref{Fig1} for $\gamma=1$ and $s=\lambda$, where the parity of the bonds has been made explicit.

By exponentially tilting the previous generator one can uncover the
fluctuation relation in the modified dynamics. Indeed, one has explicitly {\small
\begin{align*}
\mathcal{L}_\lambda^{\mrm{Doob}}(s) &= \gamma_1 \mathrm{e}^{-(s+\lambda)} \alpha(s) \sum_{x\, \mrm{even}} \ketbra{x+1}{x} + \nonumber\\ 
& + \gamma_2 \mathrm{e}^{-(s+\lambda)} \alpha(s)^{-1} \sum_{x\, \mrm{odd}} \ketbra{x-1}{x} + \nonumber\\ 
& + \gamma_1 \mathrm{e}^{s+\lambda} \alpha(s)^{-1} \sum_{x\, \mrm{odd}} \ketbra{x+1}{x} +\nonumber\\ 
& + \gamma_2 \mathrm{e}^{s+\lambda} \alpha(s) \sum_{x\, \mrm{even}} \ketbra{x-1}{x} + \nonumber\\
&- \sqrt{\gamma_1^2 + \gamma_2^2 + 2\gamma_1 \gamma_2 \cosh(2s)}\mathbbm{1},
\end{align*} }
so that at the level of the generator it turns out that
\begin{equation}
\mathcal{L}_\lambda^{\mrm{Doob}}(s) = A_s \mathcal{L}_{-\lambda-2s}^{\mrm{Doob}}(s) A_s^{-1},
\end{equation}
where the transformation $A_s(\cdot)A_s^{-1}$ exchanges the terms $ \alpha(s) \sum_{x\, \mrm{even}} \ketbra{x+1}{x} $ and $\alpha(s)^{-1} \sum_{x\, \mrm{odd}} \ketbra{x+1}{x}$ and preserves the spectrum. Therefore, the symmetry on the scaled cumulant generating function reads
\begin{equation}
\theta^{\mrm{Doob}}(\lambda) = \theta^{\mrm{Doob}}(-\lambda-2s).
\end{equation}
This can indeed be easily verified from the explicit expression of $\theta^{\mrm{Doob}}(\lambda) $
\begin{align}
\theta^{\mrm{Doob}}(\lambda) &= \sqrt{\gamma_1^2 + \gamma_2^2 + 2\gamma_1 \gamma_2 \cosh(2s+2\lambda)} \nonumber \\
&- \sqrt{\gamma_1^2 + \gamma_2^2 + 2\gamma_1 \gamma_2 \cosh(2s)}.
\end{align}

\subsection{Two-dimensional observable for a totally asymmetric random walk}
\begin{figure}[t]
\centering
\includegraphics[scale=0.27]{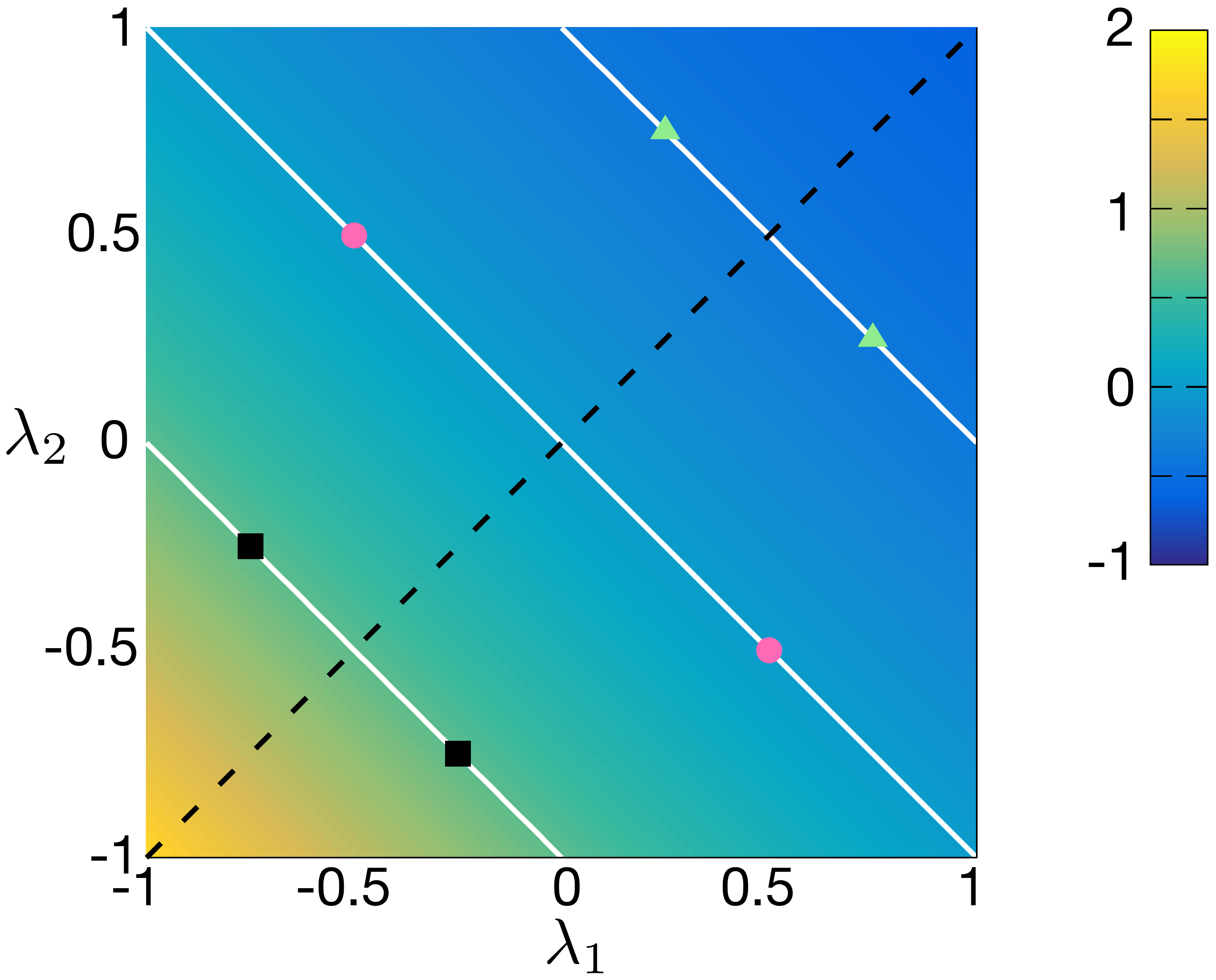}
\caption{{\bf Totally asymmetric random walk}. Scaled cumulant generating function $\theta(\underline{\lambda})$ given by \eqref{thetaex2}. The symmetry $\theta(\underline{\lambda})=\theta(U\cdot\underline{\lambda})$ can be observed by looking at some points ($\lambda_1,\lambda_2$) and their symmetric pairs (given by symbols) with respect to $\lambda_1=\lambda_2$ (black dashed line). We further notice that $\theta(\underline{\lambda})$ is invariant for all $\lambda_1+\lambda_2=\text{constant}$: e.g. $\lambda_1+\lambda_2=-1,0,1$ (white solid lines). }
\label{Fig2}
\end{figure}

Consider a totally asymmetric random walk, namely a particle hopping clockwise with rate $\gamma$ on a ring of $L$ sites. As in the previous example, the configuration is completely specified at any time by the position of the particle in the lattice. Let us consider now a vectorial observable
\begin{equation}
\underline{K}(\omega_t)= \begin{pmatrix}
    K_{\mrm{even}}(\omega_t) \\
    K_{\mrm{odd}}(\omega_t)
   \end{pmatrix},
\end{equation}
where $K_{\mrm{odd}}$ ($K_{\mrm{even}}$) is the number of jumps from odd (even) sites, and a transformation $\mathcal{R}$ acting on the configurations that translates the position in the lattice by an odd number of sites. This transformation at the trajectory level induces a transformation of the observable described by a matrix $U$. Explicitly, since we are exchanging even and odd sites, the map can be written as follows
\begin{equation}
U=\begin{pmatrix}
  0 &1 \\
  1 &0
  \end{pmatrix}, \quad  \underline{K}(\mathcal{R}\omega_t) = U\cdot \underline{K}(\omega_t).
\end{equation}
The exponentially tilted (relatively to the observable $\underline{K}$) generator of the stochastic process reads
\begin{equation}\label{tiltex2}
\mathcal{L}_{\underline{\lambda}}= \gamma \mathrm{e}^{-\lambda_1} \sum_{x \,\mrm{even}} \ketbra{x+1}{x}  + \gamma \mathrm{e}^{-\lambda_2} \sum_{x \,\mrm{odd}} \ketbra{x+1}{x} - \gamma \mathbbm{1}.
\end{equation}
In this case, negative $\lambda_1$ ($\lambda_2$) enhance jumps starting from even (odd) sites. and positive values of the biasing field do the opposite. The tilted generator can be diagonalized exactly. 
Indeed, by assuming the following ansatz for the right eigenvector $\ket{r_{q}(\underline{\lambda})}$ corresponding to the eigenvalue $\xi_q(\underline{\lambda})$
\begin{equation}
\ket{r_{q}(\underline{\lambda})} = \sum_{m=0}^{L-1} c_m(\underline{\lambda}) \mathrm{e}^{imq} \ket{m},
\end{equation} 
one arrives at a system of coupled linear equations for the coefficients $c_m(\underline{\lambda})$. In particular, for $m \in \{1, \ldots, L/2\}$ one has two different sets of equations corresponding to even and odd jumps
\begin{equation}\label{coeffex2}
  \begin{cases}
    c_{2m}(\underline{\lambda}) \big(\xi_q(\underline{\lambda}) + \gamma \big) - \gamma \mathrm{e}^{-\lambda_2}\mathrm{e}^{-iq} c_{2m-1}(\lambda) =0,\\
    c_{2m-1}(\underline{\lambda}) \big(\xi_q(\underline{\lambda}) + \gamma \big) - \gamma \mathrm{e}^{-\lambda_1}\mathrm{e}^{-iq} c_{2m-2}(\lambda) =0,
  \end{cases}
\end{equation}
that in turn result into
\begin{equation}
c_{2m}(\underline{\lambda}) \, \big(\xi_q(\underline{\lambda}) + \gamma\big)\!^2 - c_{2m-2}(\underline{\lambda}) \, \gamma^2 \mathrm{e}^{-(\lambda_1+\lambda_2)}\mathrm{e}^{-i2q} =0.
\end{equation}
By summing over $m$ and exploiting the periodic boundary conditions, one arrives at
\begin{equation}
\sum_{m=1}^{L/2} c_{2m}(\underline{\lambda}) \Big[ \big(\xi_q(\underline{\lambda}) + \gamma\big)\!^2 - \gamma^2 \mathrm{e}^{-(\lambda_1+\lambda_2)}\mathrm{e}^{-i2q} \Big] =0.
\end{equation}
Assuming $\sum_{m=1}^{L/2} c_{2m}(\underline{\lambda})\neq 0$ (we have checked this for consistency a posteriori) it turns out that the eigenvalues are
\begin{equation}
\xi_q(\underline{\lambda}) = \gamma \left( \mathrm{e}^{-\frac{\lambda_1+\lambda_2}{2}-iq} -1 \right),
\end{equation}
with $q= \frac{2\pi n}{L}$ and $n$ taking values in $\{0,1, \ldots ,L-1 \}$. Therefore one can access the scaled cumulant generating function that is
\begin{equation}\label{thetaex2}
\theta(\underline{\lambda})= \gamma \left( \mathrm{e}^{-\frac{\lambda_1+\lambda_2}{2}} -1 \right)
\end{equation}
and check that indeed it satisfies the fluctuation relation
\begin{equation}
\theta(\underline{\lambda})= \theta[(U^{-1})^T \cdot \underline{\lambda}],
\end{equation}
since $(U^{-1})^T = U$ and it just consists in exchanging $\lambda_1$ and $\lambda_2$. Actually, all the points in the $\lambda_1,\lambda_2$ plane such that $\lambda_1+\lambda_2=c$, with constant $c$, have the same value of $\theta(\underline{\lambda})$. We show this in Fig.~\ref{Fig2}, where some of the isolines have been displayed.

By means of equations \eqref{coeffex2} one can explicitly compute the right eigenvectors of the tilted generator.
In particular one finds that $c_{2m-1}= r$ and $c_{2m} = r \, \mathrm{e}^{\frac{\lambda_1-\lambda_2}{2}}$ for some normalization constant $r$. Correspondingly, the left eigenvectors are represented as follows
\begin{equation}
\bra{\ell_q (\underline{\lambda})} = \sum_{n=0}^{L-1} \mathrm{e}^{-inq} \tilde{c}_n(\underline{\lambda}) \bra{n} ,
\end{equation}
where the coefficients read $\tilde{c}_{2m-1}= \ell$ and $\tilde{c}_{2m}= \ell \, \mathrm{e}^{\frac{\lambda_2-\lambda_1}{2}} $ with normalization constant $\ell$. The orthonormality condition $\braket{\ell_{q'}}{r_q}= \delta_{q q'}$ fixes the product $r \, \ell$ to be $1/L$. The further condition $\braket{-}{r_0(\underline{\lambda})}=1$ can be used to fix $r$ and $\ell$ separately. In particular, one finds that
\begin{equation}
r= \frac{2}{L}\frac{1}{1+\mathrm{e}^{\frac{\lambda_2-\lambda_1}{2}}}.
\end{equation}

We can compute the moment generating function $Z(\underline{\lambda})$ by means of the relation $Z(\underline{\lambda})=\langle - | \mathrm{e}^{t\mathcal{L}_{\underline{\lambda}}} | x_0 \rangle$ and it reads
\begin{equation*}
Z(\underline{\lambda}) = 
    \begin{cases}
      \mathrm{e}^{-\gamma t} \cosh\left( \eta \right) + \mathrm{e}^{\frac{\lambda_1-\lambda_2}{2}}  \mathrm{e}^{-\gamma t} \sinh\left( \eta  \right) , \, \text{even} \, x_0,\\
      \mathrm{e}^{-\gamma t} \cosh\left( \eta  \right) + \mathrm{e}^{\frac{\lambda_2-\lambda_1}{2}}  \mathrm{e}^{-\gamma t} \sinh\left( \eta \right) , \ \text{odd} \, x_0,
    \end{cases}
\end{equation*}
where the variable $\eta$ has been defined as 
\begin{equation}
    \eta = \gamma t \, \mathrm{e}^{-\frac{\lambda_1+\lambda_2}{2}} .
\end{equation}
One can find the time-dependent Doob transform that is related to the following gauge transformation \cite{garrahan16a} {\small
\begin{align*} 
& G_{t'}= \sum_x \frac{\langle - | \mathrm{e}^{(t-t')\mathcal{L}_{\underline{\lambda}}} | x \rangle}{\langle - | \mathrm{e}^{(t-t')\mathcal{L}_{\underline{\lambda}}} | x_0 \rangle} \ketbra{x}{x} = \nonumber \\
& =\begin{cases}
     \sum_{m=1}^{L/2} \left( \ketbra{2m-1}{2m-1} + g(t,t',\underline{\lambda}) \ketbra{2m}{2m} \right), \\
     \sum_{m=1}^{L/2} \left( g^{-1}(t,t',\underline{\lambda}) \ketbra{2m-1}{2m-1} +  \ketbra{2m}{2m} \right),
  \end{cases} 
\end{align*} }
where the first line refers to odd $x_0$ and the second one to even $x_0$ and the function $g(t,t',\underline{\lambda})$ explicitly reads
\begin{equation}
     g(t,t',\underline{\lambda}) = \frac{1+\mathrm{e}^{\frac{s_2-s_1}{2}}\tanh\left[\gamma(t-t')\mathrm{e}^{-\frac{s_1+s_2}{2}}\right]}{1+\mathrm{e}^{\frac{s_1-s_2}{2}}\tanh\left[\gamma(t-t')\mathrm{e}^{-\frac{s_1+s_2}{2}}\right]}.
 \end{equation}
Given this transformation, the evolution of an odd initial configuration is given by the time-ordered exponential of the following time-dependent stochastic generator
\begin{align}
    \mathcal{L}^{\mrm{Doob}}_{t'} (\underline{s})&= \sum_{m=0}^{L/2-1} \Big( \gamma \mathrm{e}^{-s_1} g^{-1} \ketbra{2m+1}{2m} + \nonumber \\
      & +\gamma \mathrm{e}^{-s_2} g \ketbra{2m+2}{2m+1} + \frac{\partial_{t'}g}{g} \ketbra{2m}{2m} \Big) + \nonumber \\
      & -(\gamma + \partial_{t'}\log Z_{t'})\mathbbm{1} ,
\end{align}
 where the dependence on $t,t',\underline{s}$ in the function $g$ has been omitted to ease the notation.
 This generator describes a process where one has a different transition rate for even and odd jumps, so that the symmetry of the original dynamics is broken. Unfortunately, due to time ordering, the calculation of the evolution of the generic configuration $\ket{x_0}$ is too complicated. However, by looking at the explicit expressions of the functions $g(t,t',\underline{s})$ one can show that   
 for long enough times, $t \gg t' \gg 1$, the Doob generator tends to a time-independent generator of the form
 \begin{equation}
     \mathcal{L}^{\mrm{Doob}}(\underline{s}) = \gamma \mathrm{e}^{-(s_1+s_2)/2}\big( \sum_{x} \ketbra{x+1}{x} - \mathbbm{1} \big) ,
 \end{equation}
namely it describes just a rescaling of the original process and the symmetry is restored. This can be physically understood since the number of odd and even jumps in the long time limit tend to be equal irrespectively of the different jumps rates. One could have obtained the same result by directly computing the usual (time-independent) Doob transform. Indeed, by looking at the tilted generator \eqref{tiltex2} and using the largest eigenvalue $\theta(\underline{\lambda})$ \eqref{thetaex2} and the corresponding left eigenvector that has components $\ell_{2n-1}= 1$ and $\ell_{2n}= \mathrm{e}^{(\lambda_1-\lambda_2)/2}$ one gets
\begin{align}\label{DoobEx2}
&\mathcal{L}^{\mrm{Doob}}(\underline{s}) = \gamma \mathrm{e}^{-s_1} \sum_{x\, \mrm{even}} \frac{\ell_{x+1}}{\ell_x}(\underline{s}) \ketbra{x+1}{x} + \nonumber \\
&+ \gamma \mathrm{e}^{-s_2} \sum_{x\, \mrm{odd}} \frac{\ell_{x+1}}{\ell_x}(\underline{s})  \ketbra{x+1}{x} - \big(\gamma + \theta(\underline{s})\big) \mathbbm{1}\nonumber\\ 
& = \gamma \mathrm{e}^{-(s_1+s_2)/2}\big( \sum_{x} \ketbra{x+1}{x} - \mathbbm{1} \big) .
\end{align}
By means of an exponential tilting one arrives at the following scaled cumulant generating function
\begin{equation}
\theta^{\mrm{Doob}}_{\underline{\lambda}}(\underline{s}) = -\gamma \mathrm{e}^{-(s_1+s_2)/2} \left( 1- \mathrm{e}^{-(\lambda_1+\lambda_2)/2} \right).
\end{equation}
Given the parameter $\underline{\lambda}'$ defined as 
\begin{equation}
\underline{\lambda}'= (U^{-1})^T \cdot (\underline{\lambda} + \underline{s}) - \underline{s} = \begin{pmatrix}
\lambda_2 + s_2 - s_1 \\
\lambda_1 + s_1 - s_2
\end{pmatrix},
\end{equation}
one easily verifies the fluctuation relation $\theta^{\mrm{Doob}}(\underline{\lambda})= \theta^{\mrm{Doob}}(\underline{\lambda}')$. The Jarzynski-like relation \eqref{Jarz} in this case reads
\begin{equation}
\langle \mathrm{e}^{(s_1 - s_2)( K_{\mrm{even}} - K_{\mrm{odd}})} \rangle =1,
\end{equation}
which means $(s_1-s_2)\langle  K_{\mrm{even}} - K_{\mrm{odd}} \rangle \leq 0$. The interpretation of this result is quite straightforward: if $s_1>s_2$ the field suppresses the probability of even jumps with respect to odd jumps. This is confirmed by looking at the ratio between the even jump rate $\mathrm{e}^{-s_1}g^{-1}$ and the odd jump rate $\mathrm{e}^{-s_2}g$ that is always smaller than one, for $s_1>s_2$, and tends to one for long times. Indeed, one has
\begin{align}
   & \frac{\mathrm{e}^{-s_1}g^{-1}}{\mathrm{e}^{-s_2}g}= \mathrm{e}^{s_2-s_1} g^{-2}= \nonumber \\ &=\left(\frac{\mathrm{e}^{\frac{s_2-s_1}{2}}+\tanh\left[\gamma(t-t')\mathrm{e}^{-\frac{s_1+s_2}{2}}\right]}{1+\mathrm{e}^{\frac{s_2-s_1}{2}}\tanh\left[\gamma(t-t')\mathrm{e}^{-\frac{s_1+s_2}{2}}\right]}\right)^2 \leq 1,
\end{align}
because $a+b \leq 1+ab$ for any $0 \leq a\leq 1$ and $0\leq b\leq 1$.

\subsection{One-dimensional Ising model}
We now derive the fluctuation relation for a type-B observable, such as the time-integrated magnetization 
\be
M(\omega_t)=\int_0^{t}m(t') dt',\quad\text{with}\quad m(t)=\sum_{k=1}^L s_k(t)
\label{Mint}
\ee
of a one-dimensional Ising model of $L$ sites with periodic boundary conditions and undergoing  Glauber dynamics \cite{glauber63a}. As a transformation $V$ on the generator, we consider flipping all spins. The Hamiltonian of the system is $H=-J \sum_{k=1}^L s_k s_{k+1}$, with $J$ being the interaction constant and spin values $s_k\in \{-1,1\}$. Every configuration of the system $\{C\}=\{s_k\}_{k=1,...,L}$ is represented as a vector in a Hilbert space,
\be
\ket{C}=\bigotimes_{k=1}^L \begin{pmatrix}
\frac{1+s_k}{2}\\\frac{1-s_k}{2}
\end{pmatrix}\, ,
\ee
such that $s_k=1$ corresponds to $\ket{1}_k=(1,0)^T$ and $s_k=-1$ to $\ket{0}_k=(0,1)^T$.
Thus the probability of the system at time $t$ is encoded in the vector $\ket{P(t)}=\sum_{i=1}^{2^L} P(C_i,t)\ket{C_i}$, with $P(C_i,t)$ standing for the probabilities of the different configurations $C_i$ at time $t$. The evolution equation for the system with the Glauber dynamics is thus given by 
\be
\partial_t \ket{P(t)}=\mathcal{L}\ket{P(t)}\, ,
\nonumber
\ee
where \cite{glauber63a,schutz01a}
\be
\mathcal{L}=\frac{1}{2}\sum_{k=1}^L (\sigma_k^x-\mathbbm{1}) 
\left[ \mathbbm{1}-\frac{\gamma}{2}\sigma_k^z (\sigma_{k-1}^z+\sigma_{k+1}^z) \right]
\, ,
\label{genGlauber}
\ee
with $\gamma=\tanh(2 \beta J)$ and inverse temperature $\beta=1/(k_BT)$. Here $\sigma_k^{x,z}$ are the standard Pauli matrices acting on site $k$, and $\mathbbm{1}$ is the $2^L\times 2^L$ identity matrix $\mathbbm{1}=\Motimes_{k=1}^L \mathbbm{1}_k$. The Glauber generator \eqref{genGlauber} encodes the spin flip at site $k$ by means of the $\sigma_k^x$ operator at a rate given by $1/2-\gamma s_k (s_{k-1}+s_{k+1})/4$. 

\begin{figure}[t]
\centering
\includegraphics[scale=0.37]{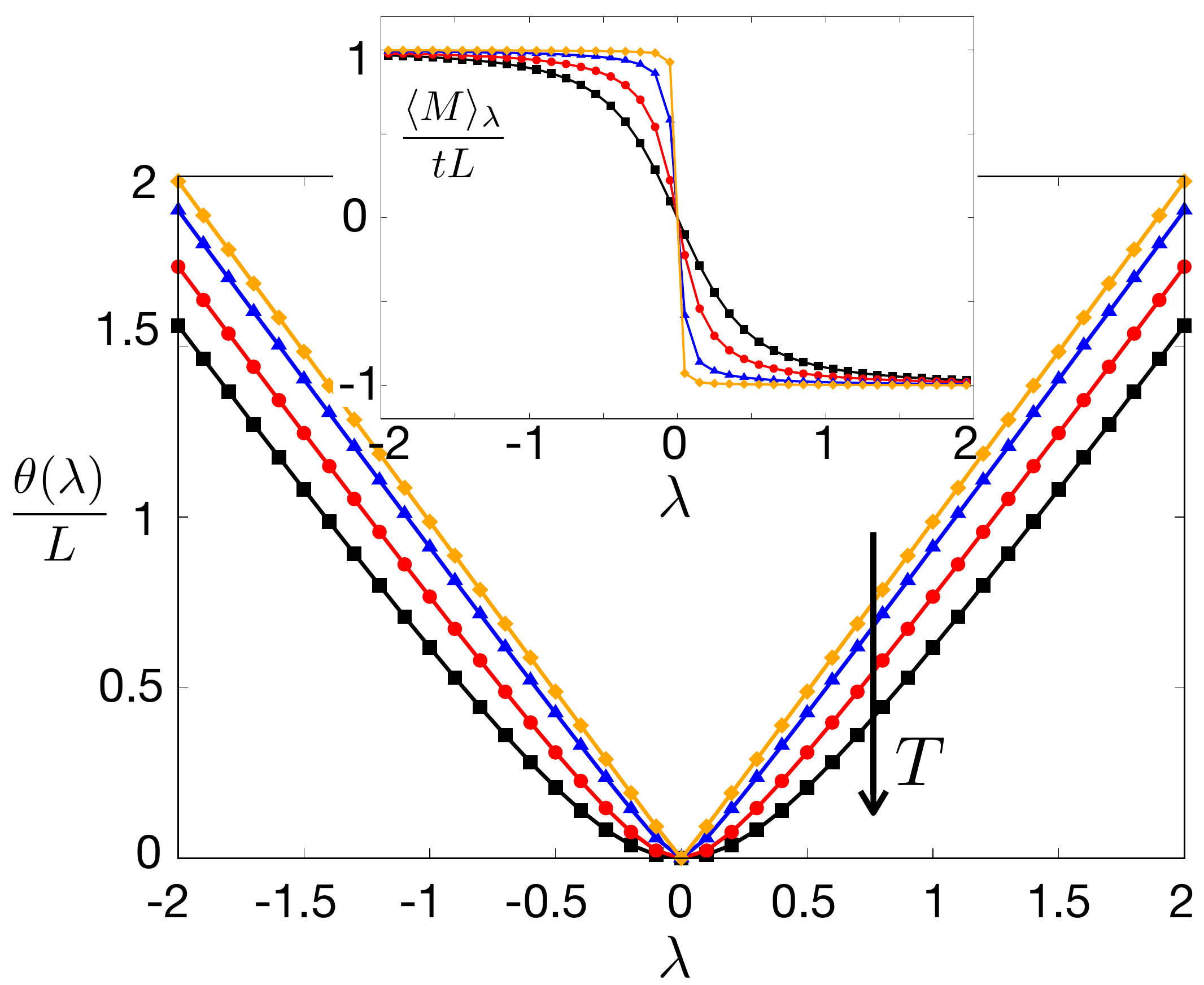}
\caption{{\bf One-dimensional Ising model}. Scaled cumulant generating function of the time-averaged magnetization $\theta(\lambda)/L$ for $L=20$ and different temperatures ($T=1,2,5,\infty$). Inset: Time-averaged magnetization per spin $\langle M \rangle_{\lambda}/(tL)$ as a function of the biasing field $\lambda$.}
\label{Fig3}
\end{figure}

In order to bias the original generator to have a given time-integrated magnetization, we firstly write the magnetization in an operatorial form as 
\be
\hat{m}=\sum_{k=1}^L \sigma_k^z\, ,
\ee
so that the tilted generator is then
\be
\mathcal{L_{\lambda}}=\sum_{k=1}^L 
\left\{
\frac{1}{2}(\sigma_k^x-\mathbbm{1}) 
\left[
\mathbbm{1}-\frac{\gamma}{2}\sigma_k^z (\sigma_{k-1}^z+\sigma_{k+1}^z)
\right]
-\lambda \sigma_k^z
\right\} 
\, .
\label{genGlaubertilt}
\ee
Notice that for negative (positive) $\lambda$ we are biasing the system towards a positive (negative) magnetization. 
At this point it is easy to check that the original generator \eqref{genGlauber} is symmetric under the spin flipping transformation, namely if we take 
$$V=\prod_{k=1}^L \sigma_k^x \, $$
we have that
\be
\mathcal{L}=V\mathcal{L}V^T\, .
\ee
On the other hand by flipping all spins the magnetization changes sign, $V\hat{m}V^T=-\hat{m}$, and so does the time-integrated one $M(\mathcal{R}\omega_t)=-M(\omega_t)$, where $\mathcal{R}$ consists in flipping all the spins of each configuration of the trajectory $\omega_t$. We thus have that $U=-1$ and obtain (see Eq. \eqref{tilsymmrel}) 
\be
\mathcal{L}_{\lambda}=V\mathcal{L}_{-\lambda}V^T\, ,
\ee
which is easy to verify. As a consequence the scaled cumulant generating function has the following symmetry: $\theta(\lambda)=\theta(-\lambda)$. We have checked this symmetry computing $\theta(\lambda)$ by numerical exact diagonalization of the tilted generator \eqref{genGlaubertilt}. Results are shown in Fig.~\ref{Fig3} for a system with $L=20$ sites and different temperatures (considering $J=1$ and $k_B=1$). We have as well represented the time-averaged magnetization per spin $\langle M(\omega_t) \rangle_{\lambda}/(Lt)=-\theta'(\lambda)/L$ for different values of the biased (see inset to Fig.~\ref{Fig3}). 

As the symmetry $\theta(\lambda)=\theta(-\lambda)$ holds, we can state in virtue of \eqref{FRlong}, that in the presence of a field $E$ the FR 
$$\theta_E(\lambda)=\theta_E(-\lambda-2E) \, $$
holds alike.

\subsection{One-dimensional three-state Potts model}
Our last example concerns the derivation of the fluctuation relation in the one-dimensional three-state Potts model. In this case we will see that the transformation leaving invariant the dynamics corresponds to a spin rotation by an angle of $\pm 2\pi/3$ rad. Hence, considering again the time-integrated magnetization---which is a bidimensional observable---we derive a fluctuation relation connecting different rotated magnetizations.

The three-state Potts model \cite{potts52a} consists of a spin system with $L$ spins that are in one of the three states $s_k\in\{0,1,2\}$ with $k=1,...,L$, which are uniformly distributed about the circle at angles $\theta_k=2\pi s_k/3$. We are considering the one-dimensional system with periodic boundary conditions, whose Hamiltonian is given by
$$H=-J\sum_{k=1}^L \delta(s_k,s_{k+1})$$ 
where $\delta(i,j)$ is the Kronecker delta $\delta_{ij}$. Every configuration of the system $\{C\}=\{s_k\}_{k=1,...,L}$ is represented as a vector in a Hilbert space,
\be
\ket{C}=\bigotimes_{k=1}^L \left[ \delta(s_k,0),\delta(s_k,1),\delta(s_k,2) \right]^T\, ,
\ee
such that $s_k=0$, $s_k=1$ and $s_k=2$ correspond to $\ket{0}_k=(1,0,0)^T$, $\ket{1}_k=(0,1,0)^T$ and $\ket{2}_k=(0,0,1)^T$ respectively.
Thus the probability of the system at time $t$ is encoded in the vector $\ket{P(t)}=\sum_{i=1}^{3^L} P(C_i,t)\ket{C_i}$, with $P(C_i,t)$ standing for the probabilities of the different configurations $C_i$ at time $t$. The evolution equation for the system is thus given by 
\be
\partial_t \ket{P(t)}=\mathcal{L}\ket{P(t)}\, .
\nonumber
\ee
The generator of the Glauber dynamics in this case can be conveniently written as follows (see Appendix \ref{app: generator})
\begin{equation}
    \mathcal{L}= \sum_{k=1}^L ( \mathcal{L}_k^{01} + \mathcal{L}_k^{02} + \mathcal{L}_k^{12} )
\end{equation}
where the index $k$ identifies at which site the transition occurs and the superscript $(ij)$ specifies the transition $\ket{i}\leftrightarrow \ket{j}$. One can show that $\mathcal{L}_k^{ij}$ reads (see Appendix \ref{app: generator} for more details)
\begin{align}
    &\mathcal{L}_k^{ij} = \Big( \ketbra{i}{j}_k + \ketbra{j}{i}_k - \ketbra{i}{i}_k - \ketbra{j}{j}_k \Big)\times \nonumber \\
    &\times \frac{1}{2} \Big( \mathbbm{1} - \Gamma^{ij}_k \big( \gamma \mathbbm{1} + \frac{\delta}{2}\Gamma^{ij}_{k-1}\Gamma^{ij}_{k+1} \big) \big(\Gamma^{ij}_{k-1}+ \Gamma^{ij}_{k+1} \big)\Big)
    \label{PottsGen}
\end{align}
with $\gamma = \tanh(\beta J/2)$, $\delta = \tanh(\beta J) - 2 \tanh(\beta J/2)$ and 
\begin{equation}\nonumber
     \Gamma^{01}_k =\!\!\begin{pmatrix}
                1 & 0 &0 \\
                0 & -1 & 0 \\
                0 & 0 & 0
               \end{pmatrix},
   \Gamma^{02}_k =\!\!\begin{pmatrix}
                1 & 0 &0 \\
                0 & 0 & 0 \\
                0 & 0 & -1
               \end{pmatrix},
   \Gamma^{12}_k =\!\!\begin{pmatrix}
                0 & 0 &0 \\
                0 & 1 & 0 \\
                0 & 0 & -1
               \end{pmatrix}\!\!.
\end{equation}
The time-integrated magnetization reads
\be
\underline{M}(\omega_t)=\int_0^{t}\underline{m}(t') dt',
\label{Mvecint}
\ee
with $\underline{m}(t)=\sum_{k=1}^L( \cos 2\pi s_k(t)/3,  \sin 2\pi s_k(t)/3)^T$,
which in operatorial form can be written as follows
\be
\hat{\underline{m}}=\sum_{k=1}^L \underline{\hat{m}}_k=  \sum_{k=1}^L \left(\frac{1}{2}(\Gamma_k^{01} + \Gamma_k^{02}),\frac{\sqrt{3}}{2} \Gamma_k^{12}\right)^T\, .
\ee
\begin{figure}[t]
\centering
\includegraphics[scale=0.40]{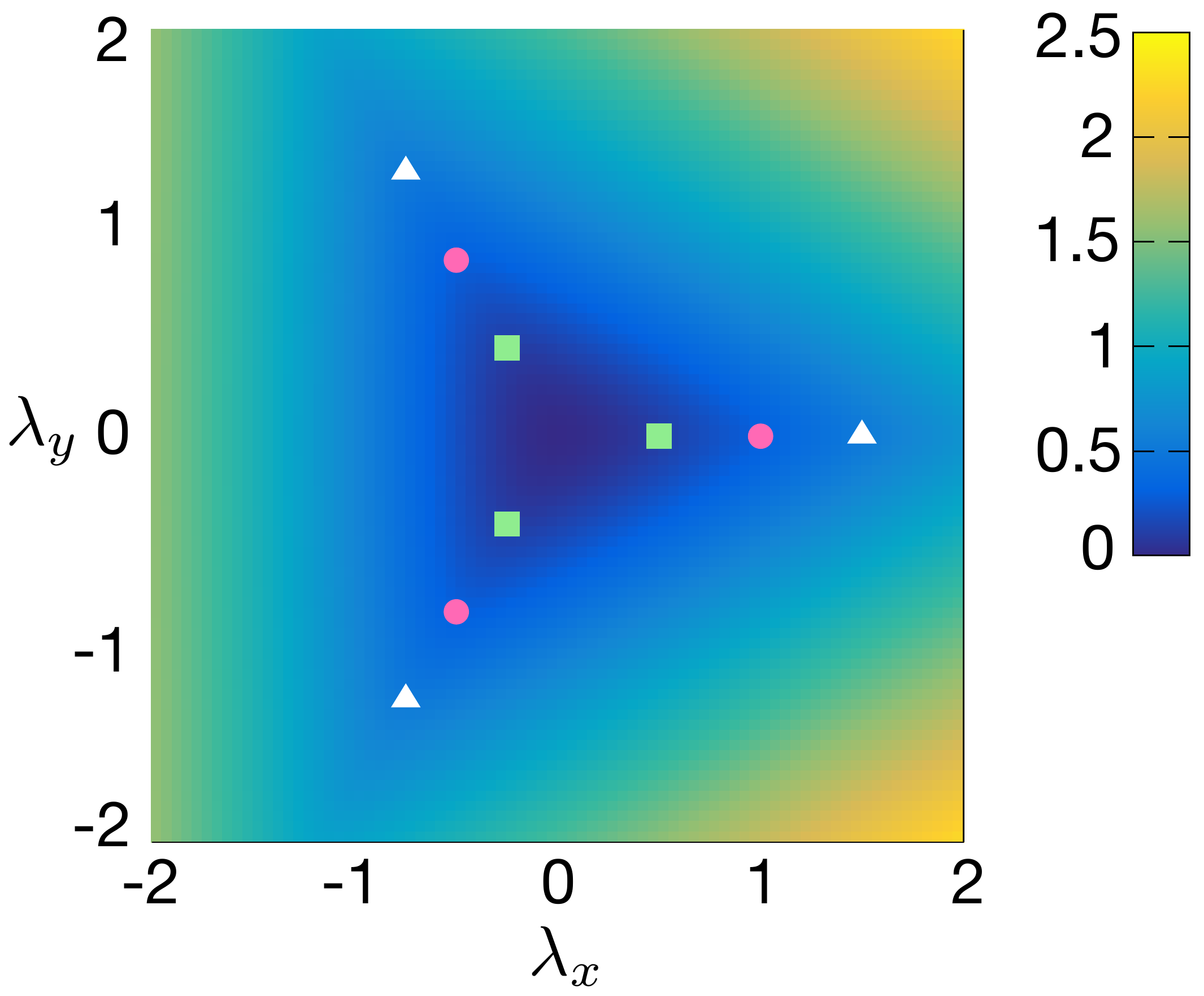}
\caption{{\bf One-dimensional three-state Potts model}. Scaled cumulant generating function of the time-averaged magnetization $\theta(\underline{\lambda})/L$ for $L=10$ and $T=2$. Different symbols represent points with the same value of $\theta(\underline{\lambda})$, and are related by a rotation of $\pm2\pi/3$ rad, hence $\theta(\underline{\lambda})=\theta(U_{\pm\frac{2\pi}{3}}\cdot\underline{\lambda})$ for $|\underline{\lambda}|=0.5, 1, 1.5$.}
\label{Fig4}
\end{figure} 
Thus the tilted generator reads
\begin{equation}
    \mathcal{L}_{\underline{\lambda}}= \sum_{k=1}^L ( \mathcal{L}_k^{01} + \mathcal{L}_k^{02} + \mathcal{L}_k^{12} - \underline{\lambda}^T\cdot \underline{\hat{m}}_k)\, .
    \label{PottsGentilt}
\end{equation}
It is easy to check that the transformation that leaves invariant the original generator is the rotation by $2\pi/3$ rad, given by $V_{\frac{2\pi}{3}}=\prod_{k=1}^L (\ketbra{1}{0}_k+\ketbra{2}{1}_k+\ketbra{0}{2}_k)$, and the rotation by $-2\pi/3$ rad, given by $V_{-\frac{2\pi}{3}}=V_{\frac{2\pi}{3}}^T$. We thus have 
\be
    \mathcal{L}=V_{\pm \frac{2\pi}{3}}    \mathcal{L}V_{\pm \frac{2\pi}{3}}^T \,.
\ee
Further, it is straightforward to verify that $\underline{M}(\mathcal{R}_{\pm}\omega_t)=U_{{\pm}\frac{2\pi}{3}}\cdot \underline{M}(\omega_t)$, with $\mathcal{R}_{\pm}\omega_t$ being a trajectory with all its configurations rotated by ${\pm}2\pi/3$, and $U_{\pm\frac{2\pi}{3}}$ the rotation matrix of $\pm\frac{2\pi}{3}$ rad, given by
\be
U_{\pm\frac{2\pi}{3}}=\begin{pmatrix}
                \cos(2\pi/3) & \pm\sin(2\pi/3) \\
                \mp\sin(2\pi/3) & \cos(2\pi/3) 
               \end{pmatrix}.
               \nonumber
\ee
The tilted generator then has the symmetry 
$$\mathcal{L}_{\underline{\lambda}}=V_{\pm \frac{2\pi}{3}}    \mathcal{L}_{U_{\pm\frac{2\pi}{3}}\cdot \underline{\lambda}} V_{\pm \frac{2\pi}{3}}^T \, $$ 
which leads to the following symmetry of the scaled cumulant generating function: 
$$\theta(\underline{\lambda})=\theta(U_{\pm\frac{2\pi}{3}}\cdot\underline{\lambda}) .$$ 
We have numerically checked this symmetry by diagonalizing \eqref{PottsGentilt} for $L=10$ spins, obtaining the largest eigenvalue $\theta(\underline{\lambda})$ presented in Fig.\ref{Fig4}. There we have highlighted the points having the same $\theta(\underline{\lambda})$, which are those related by a rotation of $\pm \frac{2\pi}{3}$ rad. As a consequence of the symmetry displayed in Fig.\ref{Fig4}, we have that for any constant field $\underline{E}$, the fluctuation relation 
$$\theta_{\underline{E}}(\underline{\lambda})=\theta_{\underline{E}}[U_{\pm\frac{2\pi}{3}}\cdot(\underline{\lambda}+\underline{E})-\underline{E}]$$
holds, as was shown in Eq. \eqref{FRlongt}. 


\section{Conclusion}
\label{sec: conclusion}
In this work we have clarified the conditions that allow to prove symmetry-induced fluctuation relations in the context of classical continuous-time Markov chains. In particular, we proved that given a dynamics with a certain symmetry and choosing a suitable observable, it is always possible to find a related dynamics where the symmetry is explicitly broken but persists at the level of the observable's fluctuations. The new dynamics is obtained from the original one via a generalized Doob transform. This approach leads to FRs for observables that are not necessarily time-antisymmetric, in contrast to the usual FRs for current-like quantities. 
Focusing on trajectory transformations that act uniformly at the level of each configuration, we have shown how a suitable high-dimensional observable can always be found in order to satisfy the symmetry-induced fluctuation relation. While this proof guarantees the existence of such observables, in practice in many cases it is possible to find many other suitable low-dimensional observables of interest as well and we suggest a systematic way to build them. We also provided an alternative proof of these FRs for long times looking at the symmetry properties of the generator.  

We illustrated our general results with four different examples to highlight the presence of the fluctuation relation in some unexpected contexts, namely for time-symmetric observables, like activity related quantities and time-integrated magnetizations. In particular, the first two examples dealt with a particle hopping on a ring, where we discuss a one-dimensional observable that can have both positive and negative values and a two-dimensional observable that only has positive values. In the third and fourth examples, we discussed the integrated magnetization in the Glauber-Ising and in the Glauber-3-state-Potts models, respectively, highlighting the symmetry properties of the scaled cumulant generating function.

Here we focussed on systems with continuous-time Markov dynamics and for concreteness on time-local trajectory transformations. We anticipate further developments: (i) for classical systems, generalisation to discrete Markov chains is straightforward; (ii) it should also be possible to extend our results to open quantum systems described either by discrete-time or continuous-time quantum Markov chains; (iii) it will be interesting to try to uncover novel FRs emerging from trajectory-space symmetries whose transformations are not local in time. These issues will be the subject of future investigations.

\section*{Acknowledgments}
The authors thank F. Carollo for stimulating discussions. The research leading to these results has received funding from the European Union’s Horizon 2020 research and innovation programme under the Marie Sklodowska-Curie Cofund Programme Athenea3I Grant Agreement No. 754446, from the European Regional Development Fund, Junta de Andaluc\'ia-Consejer\'ia de Econom\'ia y Conocimiento, Ref. A-FQM-175-UGR18, and from the EPSRC Grant No. EP/R04421X/1. We acknowledge the use of Athena at HPC Midlands+, which was funded by the EPSRC on grant EP/P020232/1, in this research, as part of the HPC Midlands+ consortium. We are also grateful for the computational resources and assistance provided by PROTEUS, the supercomputing center of the Institute Carlos I for Theoretical and Computational Physics at the University of Granada, Spain.

\appendix 

\section{Low dimensional observables}
\label{app: observables}

In the following we present a systematic way to find low dimensional observables satisfying the assumption number $2$. For concreteness, let us focus on the $4$-sites totally asymmetric random walk already introduced at the end of Section \ref{subsec: obs} with a map in configuration space that is $Rx= x+1(\text{MOD}\, 4)$.
Already with this simple model, one can note that finding low dimensional observables such that $\underline{K}(\mathcal{R}\omega_t)= U \underline{K}(\omega_t)$ is a non-trivial task.
Indeed, if we concentrate for instance on scalar observables, we immediately see that there are cases in which is not possible to find the matrix $U$ (just a number for scalar observables) independent of $\omega_t$. Explicitly, we can choose $K_1(\omega_t)=Q_{1\to 2}(\omega_t) $ so that $K_1(\mathcal{R}\omega_t)= Q_{4\to 1}(\omega_t) $ (see Eq. \eqref{qRtilde}). In this case a counterexample to the existence of $U$ is easily found
$$
\omega^1_t: 1 \to 2 \to 3 \,(K_1=1), \quad \mathcal{R}\omega^1_t: 2 \to 3 \to 4 \,(K_1=0)  ,
$$
while
$$
\omega^2_t: 4 \to 1 \to 2 \,(K_1=1), \quad \mathcal{R}\omega^2_t: 1 \to 2 \to 3 \,(K_1=1).
$$
A way to find low-dimensional observables such that $U$ exists is therefore a relevant problem.
The starting point is to write a vectorial observable (with maximal dimension) where each component corresponds to the number of a specific type of jump $Q_{x \to y}$. Then, we recognize the fact that the tranformation $\mathcal{R}$ acts as a permutation of the entries so that the matrix $U$ always exists and has the form of a permutation matrix (each row and each column has a single element equal to 1 and 0 in the remaining entries)
\begin{align}\label{fundK}
&\underline{K} (\mathcal{R}\omega_t) = \begin{pmatrix}
Q_{1\to 2} (\mathcal{R}\omega_t)\\
Q_{2\to 3} (\mathcal{R}\omega_t)\\
Q_{3\to 4} (\mathcal{R}\omega_t)\\
Q_{4\to 1} (\mathcal{R}\omega_t)
\end{pmatrix}
= 
\begin{pmatrix}
Q_{4\to 1} (\omega_t)\\
Q_{1\to 2} (\omega_t)\\
Q_{2\to 3} (\omega_t)\\
Q_{3\to 4} (\omega_t)
\end{pmatrix} = 
U \underline{K}(\omega_t)  ,  \nonumber\\
&\text{with} \quad 
U=
\begin{pmatrix}
0 &0 &0 &1\\
1 &0 &0 &0 \\
0 &1 &0 &0 \\
0 &0 &1 &0
\end{pmatrix}.
\end{align}
This matrix $U$ has the property $U^4=\mathbbm{1}_4$ and is therefore a representation of the cyclic group $C_4$ on a four-dimensional vector space. This is consistent with the symmetry of the problem. Note however that in order to satisfy the assumption number $2$ we just partially exploit  the symmetry of the dynamics, requiring that the transformation $R$ preserves the set of allowed jumps. The full dynamical symmetry as encoded in the transition rates (pertaining to the assumption $1$ in the Section \ref{sec: IsoFluc} of the manuscript) does not enter at this point. In particular, the transition rates of the four jumps could be all different, spoiling the assumption $1$ but not the assumption $2$.  \\
Given the ``fundamental observable'' \eqref{fundK}, one can easily find other observables  with maximal dimension $\widetilde{\underline{K}}=P\underline{K}$ by considering an invertible $4\times 4$ matrix $P$. Indeed, one can easily see that $\underline{K}(\mathcal{R}\omega_t) = P^{-1}P U P^{-1} P \underline{K}(\omega_t)$, so that
\begin{equation}
\widetilde{\underline{K}}(\mathcal{R}\omega_t)  = \overline{U} \, \widetilde{\underline{K}}(\omega_t), \quad \text{where} \quad \overline{U}= P U P^{-1}.
\end{equation} 
In order to find low dimensional observables we can think of diagonalizing the matrix $U$.
Indeed, if the matrix $P$ is such that it diagonalizes $U$, namely we find a diagonal $\overline{U}$ with the eigenvalues of $U$ as elements, the different components of $\widetilde{\underline{K}}$ transform independently and can be used as low dimensional observables. In particular, the eigenvalues of $U$ in \eqref{fundK} are $\{1,-1,i,-i\}$ (these are also the irreducible representations of the cyclic group $C_4$ on the field $\mathbb{C}$). The one-dimensional eigenvectors corresponding to the eigenvalues $1$ and $-1$ can be used as proper one-dimensional observables
\begin{align*}
&\widetilde{K}^+ = \alpha ( Q_{1\to 2} + Q_{2\to 3} + Q_{3\to 4} + Q_{4\to 1}) ,  \\
&\widetilde{K}^- = \alpha ( Q_{1\to 2} - Q_{2\to 3} + Q_{3\to 4} - Q_{4\to 1}),
\end{align*}
for some real parameter $\alpha$.
In particular, we recover the dynamical activity $\widetilde{K}^+$ and the difference between the number of even and odd jumps $\widetilde{K}^-$.
The two imaginary eigenvalues and the corresponding eigenvectors 
\begin{align*}
&\widetilde{K}^{i} = \alpha ( Q_{1\to 2} -i Q_{2\to 3} - Q_{3\to 4} +i Q_{4\to 1}) ,  \\
&\widetilde{K}^{-i} = \alpha ( Q_{1\to 2} +i Q_{2\to 3} - Q_{3\to 4} -i Q_{4\to 1}),
\end{align*}
cannot be used directly because we are interested in real observables. However, we can build a real two-dimensional observable out of them. Indeed, one has
\begin{equation}
\widetilde{\underline{K}}_2 : = 
\begin{pmatrix}
\widetilde{K}^{i}  \\
\widetilde{K}^{-i} 
\end{pmatrix}, \quad 
\widetilde{\underline{K}}_2(\mathcal{R}\omega_t) = 
\begin{pmatrix}
i &0 \\
0 &-i
\end{pmatrix}
\widetilde{\underline{K}}_2 (\omega_t),
\end{equation}
so that by means of a similarity transformation $V$ one can define $\underline{K}_2 := V \widetilde{\underline{K}}_2 $ and eventually it turns out that
\begin{equation}
\underline{K}_2 (\mathcal{R}\omega_t)  = 
\begin{pmatrix}
0 &-1 \\
1 &0
\end{pmatrix} \underline{K}_2(\omega_t), 
\end{equation}
where 
\begin{equation*}
V=
\begin{pmatrix}
\frac{1}{2} & \,\frac{1}{2} \vspace{0.2cm}\\
\frac{1}{2i} &-\frac{1}{2i}
\end{pmatrix}, \qquad 
V  \begin{pmatrix}
i &0 \\
0 &-i
\end{pmatrix} V^{-1}=
\begin{pmatrix}
0 &-1 \\
1 &0
\end{pmatrix}.
\end{equation*}
A similar two-dimensional observable can also be costructed combining $\widetilde{K}^+$ and $\widetilde{K}^+$ in order to obtainthe quantity discussed in the second example of Section \ref{sec: examples}, namely the number of odd jumps and the number of even jumps considered separately.

\section{Generator of the Potts model}
\label{app: generator}

In the following we discuss in more detail a convenient way to write the generator of the Glauber dynamics for the three-state Potts model as already presented in the main text. We found it in order to ease the numerical diagonalization, taking inspiration from the simpler case of the Ising model.
First of all, we separate the contribution given by the different transitions occurring at each site by writing the generator as
\begin{equation}
    \mathcal{L}= \sum_{k=1}^L ( \mathcal{L}_k^{01} + \mathcal{L}_k^{02} + \mathcal{L}_k^{12} ),
\end{equation}
where the index $k$ identifies at which site the transition occurs and the superscript specifies the kind of transition. 

Let us start analyzing the term $\mathcal{L}_k^{01}$. By fixing the transition to be $0 \to 1$ on the site $k$ one has to specify the nearest neighbours in order to determine the rate. Indeed, given the Hamiltonian $H=-J\sum_{k=1}^L \delta(s_k,s_{k+1})$, one has the following 9 possibilities for $\Delta E= E_{\text{final}}-E_{\text{initial}}$
\begin{align}
  &000 \to 010 \quad  \Delta E = 2J ,\nonumber\\ 
  &100 \to 110 \quad  \Delta E = 0  ,\nonumber\\ 
  &001 \to 011 \quad  \Delta E = 0  ,\nonumber\\ 
  &101 \to 111 \quad  \Delta E = -2J ,\nonumber \\
  &002 \to 012 \quad  \Delta E = J ,\nonumber\\ 
  &200 \to 210 \quad  \Delta E = J ,\nonumber\\
  &202 \to 212 \quad  \Delta E = 0 ,\nonumber\\
  &102 \to 112 \quad  \Delta E = -J ,\nonumber\\
  &201 \to 211 \quad  \Delta E = -J .
\end{align}
The signs are reversed if instead we consider the transition $1 \to 0$ on the site $k$.
According to Glauber's recipe, these transitions correspond to five different rates determined by the energy differences $\Delta E$ as follows
\begin{equation}\label{ratesPotts1}
    W_{i \to j} = \frac{1}{1+\mathrm{e}^{\beta\Delta E_{ij}}}.
\end{equation}
With this information, one can try to write the operator $\mathcal{L}_k^{01}$ as the product of an off-diagonal term, describing the transition, multiplied by a diagonal term that produces the right transition rate depending on the nearest neighbours configuration.
In particular, one can show that the following expression for $\mathcal{L}_k^{01}$ does the job
\begin{align}
    &\mathcal{L}_k^{01} = \Big( \ketbra{0}{1}_k + \ketbra{1}{0}_k - \ketbra{0}{0}_k - \ketbra{1}{1}_k \Big)\times \nonumber \\
    &\times \frac{1}{2} \Big( \mathbbm{1} - \Gamma^{01}_k \big( \gamma \mathbbm{1} + \frac{\delta}{2}\Gamma^{01}_{k-1}\Gamma^{01}_{k+1} \big) \big(\Gamma^{01}_{k-1}+ \Gamma^{01}_{k+1} \big)\Big)
\end{align}
 where the $3\times 3$ matrix $\Gamma^{01}_k$ acts nontrivially only on the $k_{\text{th}}$ spin and reads
 \begin{equation}
     \Gamma^{01}_k =\begin{pmatrix}
                1 & 0 &0 \\
                0 & -1 & 0 \\
                0 & 0 & 0
               \end{pmatrix} .
     \end{equation}
Indeed, one can check that only five different rates are allowed by the second line of this expression and read
\begin{align}\label{ratesPotts2}
   \frac{1}{2}, \, \frac{1-2\gamma-\delta}{2}, \,  \frac{1+2\gamma+\delta}{2}, \, \frac{1-\gamma}{2}, \, \frac{1+\gamma}{2}.
\end{align}
Moreover, assuming that $\gamma$ and $\delta$ are positive parameters, one can check that the highest and lowest values, where both parameters $\gamma$ and $\delta$ appear, correspond to the transitions with energy $2J$ and $-2J$. Indeed, when the nearest neighbours are equal the product $\Gamma^{01}_{k-1}\Gamma^{01}_{k+1}$ acts as the identity and the sum $\Gamma^{01}_{k-1}+\Gamma^{01}_{k+1}$ is either $2$ or $-2$ depending on their value being $0$ or $1$. Finally, the action of $\Gamma^{01}_k$ fixes the overall sign, so that if the initial configuration is $000$ the rate is $(1-2\gamma-\delta)/2$ (lowest rate for the highest $\Delta E$). The other cases are also easily verified. If there is one spin in configuration $2$, then the term in front of $\delta$ is zero and the possible rates are $(1-\gamma)/2$ if the other neighbour is zero and $\Delta E=J$, or $(1+\gamma)/2$ if the other neighbour is one and $\Delta E=-J$. If the neighbours are both $2$ the sum $\Gamma^{01}_{k-1}+ \Gamma^{01}_{k+1}$ gives zero and the rate is $1/2$. The same is true if there are only zeros and ones but the neighbours have opposite values. Moreover, the factor $\Gamma^{01}_{k}$ multiplying the parenthesis ensures the change of sign if the transition $1 \to 0$ is considered instead of $0 \to 1$.

In order to complete the comparison of the rates quantitatively, we have to choose $\gamma$ and $\delta$ as follows 
\begin{equation}
   \gamma = \tanh\Big(\frac{\beta J}{2}\Big) ,
\end{equation}
\begin{equation}
   \delta = \tanh(\beta J) - 2 \tanh\Big(\frac{\beta J}{2}\Big),
\end{equation}
as can be easily verified comparing the rates \eqref{ratesPotts2} with the general formula \eqref{ratesPotts1}.
The other terms $\mathcal{L}_k^{02}$ and  $\mathcal{L}_k^{12}$ have analogous expression where the role of the different values $0,1,2$ is exchanged. In particular one should consider matrices $\Gamma^{02}_{k}$ and $\Gamma^{12}_{k}$ as follows
\begin{equation}
   \Gamma^{02}_k =\begin{pmatrix}
                1 & 0 &0 \\
                0 & 0 & 0 \\
                0 & 0 & -1
               \end{pmatrix} , \quad
   \Gamma^{12}_k =\begin{pmatrix}
                0 & 0 &0 \\
                0 & 1 & 0 \\
                0 & 0 & -1
               \end{pmatrix}.
\end{equation}
With these formulae one can readily perform a numerical diagonalization of the generator.

\bibliography{referencias-BibDesk-OK-v5}{}

\end{document}